\documentclass[3p]{elsarticle}

\usepackage{hyperref}
\date{}

\usepackage{units}
\usepackage{gensymb}
\usepackage[dvipsnames]{xcolor}
\usepackage{pgfplots}
\usepackage{amsmath}
\usepackage{mathtools}
\usepackage{txfonts}
\usepackage{dsfont}

\usepackage{bm}

\usepackage{caption}
\usepackage{subcaption}

\usepackage{todonotes}
\usepackage{layouts}
\usepackage{siunitx}

\usepackage[section]{placeins}
\usepackage[abbreviations]{glossaries-extra}

\usetikzlibrary{fadings}
\usetikzlibrary{patterns}
\usetikzlibrary{shadows.blur}
\usetikzlibrary{shapes}

\graphicspath{{figs/Sec_GeomFilters/}{MA_figs/base_deform/}{results/volscale/}{../figs/Sec_GeomFilters/}{../MA_figs/base_deform/}{../results/volscale/}{../}}
\newsavebox{\measurebox}

\journal{Engineering With Computers}

\DeclareUnicodeCharacter{2212}{-}
\DeclareMathOperator*{\argmin}{arg\,min}
\begin{document}

\newcommand{\tensor}[3]{\left[\textrm{{#1}}^{\textrm{{#3}}}_{\textrm{{#2}}}\right]}
\newcommand{\vekt}[1]{\mbox{$\boldsymbol{#1}$}}

\glsdisablehyper
\newabbreviation{tsne}{t-SNE}{t-Distributed Stochastic Neighbor Embedding}
\newabbreviation{ffd}{FFD}{free-form deformation}
\newabbreviation{pde}{PDE}{partial differential equation}
\newabbreviation{cad}{CAD}{computer-aided design}
\newabbreviation{sdf}{SDF}{signed-distance function}

\renewcommand*{\glslinkcheckfirsthyperhook}{%
	\ifdefstring\glslabel{tsne}%
	{\setkeys{glslink}{hyper=false}}%
	{}%
}

\begin{frontmatter}

\title{Neural Networks vs. Splines: Advances in Numerical Extruder Design}


\author[ilsbaddress]{Jaewook Lee\corref{mycorrespondingauthor}}
\cortext[mycorrespondingauthor]{Corresponding author}
\ead{jaewook.lee@tuwien.ac.at}

\author[catsaddress]{Sebastian Hube}
\ead{hube@cats.rwth-aachen.de}

\author[ilsbaddress,catsaddress]{Stefanie Elgeti}
\ead{stefanie.elgeti@tuwien.ac.at}

\address[ilsbaddress]{Institute of Lightweight Design and Structural Biomechanics (E317), TU Wien, Gumpendorfer Str. 7, A-1060 Vienna, Austria}
\address[catsaddress]{Chair for Computational Analysis of Technical Systems (CATS), RWTH Aachen University, Schinkelstr. 2, 52062 Aachen, Germany}

\begin{abstract}
We present a novel application of neural networks to design improved mixing elements for single-screw extruders.
Specifically, we propose to use neural networks in numerical shape optimization to parameterize geometries.
Geometry parameterization is crucial in enabling efficient shape optimization as it allows for optimizing complex shapes using only a few design variables.
Recent approaches often utilize \gls{cad} data in conjunction with spline-based methods where the spline's control points serve as design variables.
Consequently, these approaches rely on the same design variables as specified by the human designer.
While this choice is convenient, it either restricts the design to small modifications of given, initial design features -- effectively prohibiting topological changes -- or yields undesirably many design variables.
In this work, we step away from \gls{cad} and spline-based approaches and construct an artificial, feature-dense yet low-dimensional optimization space using a generative neural network.
Using the neural network for the geometry parameterization extends state-of-the-art methods in that the resulting design space is not restricted to user-prescribed modifications of certain basis shapes.
Instead, within the same optimization space, we can interpolate between and explore seemingly unrelated designs.
To show the performance of this new approach, we integrate the developed shape parameterization into our numerical design framework for dynamic mixing elements in plastics extrusion.
Finally, we challenge the novel method in a competitive setting against current free-form deformation-based approaches and demonstrate the method's performance even at this early stage.

\end{abstract}


\end{frontmatter}

%

\section{Introduction}
Modern numerical design is boosted by high-performance computers and the advent of neural networks.
While neural networks are well-established in fields such as image recognition, their power to further polymer processing is yet to be fully discovered.
This work attempts to contribute towards this goal.
We combine deep neural networks with established shape-optimization methods to enhance mixing in single-screw extruders via a novel numerical design.\par
In many polymer processing steps, screw-based machines play a crucial role.
Screws are, e.g.,  used as plasticators to prepare polymer melts for injection molding or in extruders in profile extrusion.
For simplicity, we will, in the remainder, summarize all such screw-based machines as \textit{extruders}.
Single-screw extruders (SSEs) are especially widespread among the many variants of extruders for their economic advantages and simple operation.
Economics also drives current attempts to further increase the throughput.
This increase is achieved using fast-rotating extruders.
However, the current SSE's poor mixing ability has limited the advances and, therefore, improving the mixing ability is a topic of research \cite{Celik2017Einfaerbung, gale2009mixing, Roland2019FEM, Sun_maddock, Campbell_sse, Marschik2018}.\par
Special focus is put on improved mixing elements that alleviate this limitation.
Approaches to improve mixing elements have been proposed based on analytical derivations, experimental, and simulation-based works.
In the following, we review recent developments in these three areas.
Subsequently, we outline relevant developments in the field of neural networks and, finally, motivate the use of neural nets in the numerical design of mixing elements. \par
Due to the high pressures and temperatures, analyzing the flow inside extruders is a difficult task.
Early studies thus focus on analytical models and geometrically simpler screw sections, e.g., the metering section \cite{boehm_81}.
Experiments complement these theoretical derivations and allow extending the analysis to more complex screw sections.
As reported by Gale, typical configurations rely on photomicrographs of the solidified melt \cite{gale2009mixing} that allow either investigating cross sections of the flow channel or the extrudate.
One example of such flow channel photomicrographs is Kim and Kwon's pioneering work on barrier screws via cold-screw extrusion \cite{Kim96a}.
Apart from investigating solidified melt streams, attempts to analyze the melt flow during the actual operation of extruders are occasionally reported, e.g., by Wong \textit{et al.} \cite{Wong2009}.
Despite the great success of such experiments, a standard limitation is their focus on a single operating condition.
In contrast, numerical analysis allows studying different designs and operating points at significantly reduced costs and, therefore, proliferates.
In the following, we give an overview of such numerical analyses.\par
One early example is Kim and Kwon's quasi-three-dimensional finite-element (FE) simulation of the striation formation, studying the influence of the barrier flight \cite{Kim96b}.
Another example is the work by Domingues \textit{et al.}, who obtain global mixing indices for dispersive and distributive mixing in both liquid-liquid and solid-liquid systems \cite{Domingues2012}.
Utilizing a two-dimensional simplification, their simulation domain extends from the hopper to the metering section, and their framework even allows for design optimization.\par
While these early works typically neglect mixing sections, studying the influence of mixers has recently become a vital research topic.
Celik \textit{et al.} use three-dimensional flow simulation coupled with a particle-tracking approach to determine the degree of mixing based on a deformation-based index \cite{Celik2017Einfaerbung}.
Another example is Marschik \textit{et al.}'s study comparing different Block-Head mixing screws in distributive and dispersive mixing \cite{Marschik2018}.
A comparable study -- focused on the mixing capabilities of different pineapple mixers -- is reported by Roland \textit{et al.} \cite{Roland2019FEM}.
Both works rely on three-dimensional non-Newtonian flow simulations.
Besides such works towards the numerical assessment of \textit{given} screw designs, numerical \textit{design} is also reported, however, partially in other fields of polymer processing.
For example, Elgeti \textit{et al.} aim for balanced dies and reduced die swell by applying shape optimization \cite{Elgeti12, Siegbert16}.
Design by optimization is also reported by Gaspar-Cunha and Covas, who alter the length of the feed and compression zones, the internal screw diameters of the feed and metering zone, the screw pitch, and the flight clearance \cite{Gaspar2001}.
Potente and T\"obben report another recent study devoted to mixing elements that develops empirical models for shearing sections' pressure-throughput and power consumption for numerical design \cite{Potente2002Maddock}.
Finally, a first approach combining the shape optimization methods inspired by \cite{Elgeti12} with a mixing-quantifying objective function to design mixing sections is reported in \cite{hube2021}.\par
However, the shape optimizations above share one commonality: They essentially only modify predefined geometry features.
This is accepted in many cases like die or mold design, where the final product's shape is close to the initial one (i.e., the shape variation is small).
However, topologically flexible shape parameterizations offer far greater optimization gains for mixing element design, because the optimal geometry might differ significantly from the initial shape.
The achievable improvements motivate research on geometry parametrization.\par
Established shape-parameterization approaches include radial basis functions (RBF) \cite{Kobbelt2004}, surface parameterizations using Bezier surfaces \cite{FARIN_bezier}, and surface splines \cite{piegl1996nurbs}.
All these methods may be understood as \textit{filters} that parameterize a geometry by a few variables at the price of a lack of local control.
The use of surface splines in shape optimizations can also be found in \cite{Elgeti12, Siegbert16}.
A similar concept to surface splines is \gls{ffd} \cite{Sederberg1986} that encapsulates the body-to-deform in a volumetric spline, which allows tailoring the spline further towards an efficient optimization.
An alternative approach that does, however, not parameterize the geometry as a filter is given using the computational grid's mesh nodes as shape parameters \cite{HOJJAT2014494}.
Fortunately, with the advent of neural networks, novel means of shape parameterizations offering outstanding flexibility emerged.
Finalizing the introduction, we will summarize the most relevant works in this field.\par
Many neural networks are essentially classifiers.
These neural networks are non-linear algorithms that are optimized, (i.e., trained), to determine -- possibly counterintuitive -- similarities and dissimilarities to discriminate between objects.
One typical use case is image recognition using red-green-blue (RGB) pixel data.
Neural networks can, however, be trained to classify features far beyond RGB-pixel values.
One example is style transfer or texture synthesis \cite{Johnson_styleTransfer}:
Instead of aiming at reproducing \textit{pixel} data, output images are generated in combination with \textit{perceptual} data.
This allows image transformations, where one image's style is transferred to the motive of another.
An extension of these ideas to three-dimensional shapes is first reported by Friedrich \textit{et al.} \cite{Honda_styleTransfer}.
Comparing different shape representations, the authors find that style transfer is applicable to shapes as well.\par
Our work is especially inspired by Liu \textit{et al.} \cite{VAE_shapeMorph}, who utilize a so-called \textit{Variational Shape Learner}, that learns a voxel representation of three-dimensional shapes.
\textit{Learning} here refers to creating a so-called \textit{latent space}, a low-dimensional, feature-rich embedding space to represent and morph between various shapes.
Even beyond simple shape interpolation, it is shown that -- using the latent representation -- geometry features can be transferred from one to another shape.
Successful learning of voxel-based shapes can also be found in \cite{3dgan, shapenets}.
In terms of shape representations, pointcloud-based approaches \cite{pointnet, spectral_gans, pc_ae17}, which utilize coordinates of three-dimensional point sets, as well as polygonal mesh-based approaches with either template meshes \cite{cvae_face18, mesh_vae} or multiple mesh planes \cite{atlasnet} are widely adopted.\par
While previously mentioned representations show that learning an embedding space of three-dimensional shapes is possible, each work lacks at least one of the following properties: water-tight surfaces, flexible output resolution, and smooth and continuous surface details.
Recent works satisfy the aforementioned properties by learning shapes represented by continuous implicit functions such as \glspl{sdf} \cite{Park_deepsdf} and binary occupancies \cite{chen2018implicit_decoder, o_net}, from which the shapes are extracted as isosurfaces.\par
We exploit the feature richness of this latent space as an aid to reduce the optimization space's dimension for the given mixing-element shape optimization.
The important novelty compared to recent spline-based filters is that the neural network finds -- possibly counterintuitive -- ways to commonly parameterize a set of significantly different shapes irrespective of user-defined design features.
This abstraction from the human designer yields low-dimensional yet far more flexible shape parameterizations, which sets the motivation for the work presented here.\par
This paper is structured as follows:
We start in Sec.~\ref{sec:geomfilters} by summarizing numerical shape optimization and splines, which leads to the concept of geometric filters.
Based on that, we explain in Sec.~\ref{sec:shppar} how neural networks can be utilized to create suitable geometry parameterizations for shape optimization.
In Sec.~\ref{sec:compSetup}, we review the utilized software components, summarize the proposed framework's building blocks, and detail the specific differences to spline-based shape optimization setups.
The results obtained from the new approach are presented in Sec.~\ref{sec:results}, including comparisons to current spline-based designs.
Finally, we discuss the results and outline further developments in Sec.~\ref{sec:outlook}.

\section{Geometric filters as a component of shape optimization frameworks}
\label{sec:geomfilters}
The following section discusses shape parameterizations as one building block of numerical shape optimization frameworks.
Therefore, we first introduce the general shape optimization problem.
After that, we recall spline-based shape parameterizations.
Based on this general introduction of shape optimization frameworks, we will continue by discussing the specific changes needed to adapt neural nets in Sec.~\ref{sec:shppar}.\par
\subsection{Building blocks of numerical shape optimization frameworks}

\begin{figure}[!h]
	\centering
	\tikzset{every picture/.style={line width=0.75pt}} 

	\begin{tikzpicture}[x=0.75pt,y=0.75pt,yscale=-1,xscale=1]

		\draw   (78.75,10.5) -- (510,10.5) -- (510,150) -- (78.75,150) -- cycle ;
		\draw   (87.25,23.03) -- (156.8,23.03) -- (156.8,44.78) -- (87.25,44.78) -- cycle ;
		\draw   (87.5,54.53) -- (157.05,54.53) -- (157.05,76.28) -- (87.5,76.28) -- cycle ;
		\draw    (120.07,45.25) -- (120.07,51.5) ;
		\draw [shift={(120.07,54.5)}, rotate = 270] [fill={rgb, 255:red, 0; green, 0; blue, 0 }  ][line width=0.08]  [draw opacity=0] (5.36,-2.57) -- (0,0) -- (5.36,2.57) -- cycle    ;
		\draw   (87.5,85.53) -- (157.05,85.53) -- (157.05,107.28) -- (87.5,107.28) -- cycle ;
		\draw    (120.07,76.25) -- (120.07,82.5) ;
		\draw [shift={(120.07,85.5)}, rotate = 270] [fill={rgb, 255:red, 0; green, 0; blue, 0 }  ][line width=0.08]  [draw opacity=0] (5.36,-2.57) -- (0,0) -- (5.36,2.57) -- cycle    ;
		\draw    (120.07,107.25) -- (120.07,113.5) ;
		\draw [shift={(120.07,116.5)}, rotate = 270] [fill={rgb, 255:red, 0; green, 0; blue, 0 }  ][line width=0.08]  [draw opacity=0] (5.36,-2.57) -- (0,0) -- (5.36,2.57) -- cycle    ;
		\draw   (87.5,116.53) -- (157.05,116.53) -- (157.05,138.28) -- (87.5,138.28) -- cycle ;
		\draw    (156.25,127.25) -- (183.68,127.25) -- (183.68,33.25) -- (159.25,33.25) ;
		\draw [shift={(156.25,33.25)}, rotate = 360] [fill={rgb, 255:red, 0; green, 0; blue, 0 }  ][line width=0.08]  [draw opacity=0] (5.36,-2.57) -- (0,0) -- (5.36,2.57) -- cycle    ;

		\draw (201,34) node [anchor=west] [inner sep=0.75pt]   [align=left] {Shape update (geometry kernel)};
		\draw (201,65) node [anchor=west] [inner sep=0.75pt]   [align=left] {Finite elements flow solution};
		\draw (201,96) node [anchor=west] [inner sep=0.75pt]   [align=left] {Objective computation (particle tracking emulator)};
		\draw (201,127) node [anchor=west] [inner sep=0.75pt]   [align=left] {Shape parameter update by optimization algorithm};
		\draw (123.19,34.53) node   [align=left] {FFD};
		\draw (120.58,65.73) node   [align=left] {FEM};
		\draw (119.08,96.23) node   [align=left] {OBJ};
		\draw (119.92,127.73) node   [align=left] {OPT};

	\end{tikzpicture}
	\caption{Building blocks of a shape optimization framework. The shape is updated by a geometry kernel such as \gls{ffd}.
	Subsequently, the flow field is computed using this updated shape and given as input to the objective calculator.
	Based on the current design variables and the design's objective value, the optimization algorithm computes optimized shape parameters and restarts the design loop until at least one termination criterion for the design loop is met.}
	\label{fig:ffd-framework}
\end{figure}
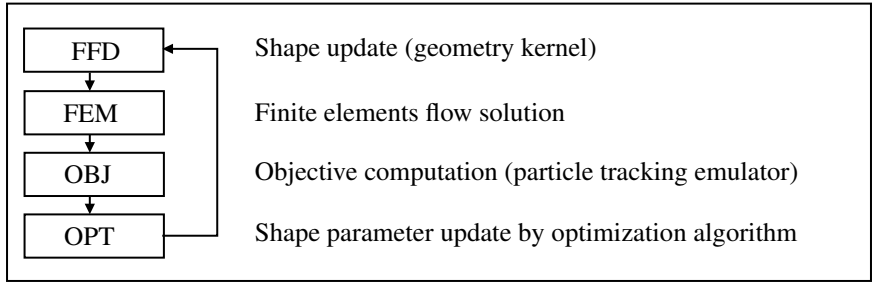

The general optimization problem is formulated as the minimization of a cost function $J$ that relates the design variables $\boldsymbol{\sigma}$ to some output -- here, the degree of mixing ability obtained with a specific mixing element, (i.e., a particular design).
In shape optimization, this minimization problem is typically solved subject to two sets of constraints:
(1) inequality and equality conditions, as well as bound constraints on the design variables and
(2) \glspl{pde} that need to be fulfilled by each design to qualify as a feasible solution.
This results in the following formulation:
\begin{subequations}\label{eq:template_opt_prob}
	\begin{alignat}{3}
		&J : \mathbb{R}^{n_{\sigma}} \mapsto \mathbb{R}\\
		\argmin_{\boldsymbol{\sigma} \in \Sigma \subset \mathbb{R}^n}\hspace{0.5em} &J\left(\boldsymbol{\sigma}\right) \label{eq:objective}\\
		\text{s.t.}\hspace{0.5em}		     &\mathbf{F}\left(\boldsymbol{\sigma}\right) = \mathbf{0} \qquad  &&\text{in} \hspace{0.5em} \Omega\left(\boldsymbol{\sigma}\right), \label{eq:pde_constraints}\\
		&\sigma_i \geq \sigma_{min,i}, &&i=1,...n_{\sigma},\label{eq:l_bounds}\\
		&\sigma_i \leq \sigma_{max,i}, &&i=1,...n_{\sigma}.\label{eq:u_bounds}
	\end{alignat}
	\label{eq:general_opt_prob}
\end{subequations}
Here, \eqref{eq:l_bounds} and \eqref{eq:u_bounds} describe bound constraints on the optimization variables $\boldsymbol{\sigma}$, whereas \eqref{eq:pde_constraints} denotes the set of governing \glspl{pde}.
One approach to numerically solve such a \textit{\gls{pde}-constraint} design problem is to alternately compute (1) shape updates and (2) the cost function value.
For the studied use case of mixing element design, this results in the computational steps depicted in Fig.~\ref{fig:ffd-framework}.
First, we update the shape (i.e., the simulation domain covering the mixing element).
We use this modified computational domain to compute the flow field from which we afterwards infer the objective (i.e., the cost function).
The design loop is closed by feeding back the cost function value to the optimization algorithm that now computes an updated shape.
This loop continues until any termination criterion such as a minimal objective decrease, a maximum number of iterations or another condition is met.\par

\subsection{Spline-based shape parameterizations}
In classical shape-optimization frameworks, the actual shape parameterization, or geometry filtering, is often achieved using splines.
The following paragraph, therefore, first provides a summary of splines illustrating how one achieves the filtering.
For a detailed description of B-splines, we refer the reader to the book of Piegl and Tiller \cite{piegl1996nurbs}.
After that, we detail on \textit{boundary splines} and \gls{ffd} as two particular use cases of spline parameterizations.\par
\begin{figure}[!h]
	\centering
	\includegraphics[width=0.5\textwidth]{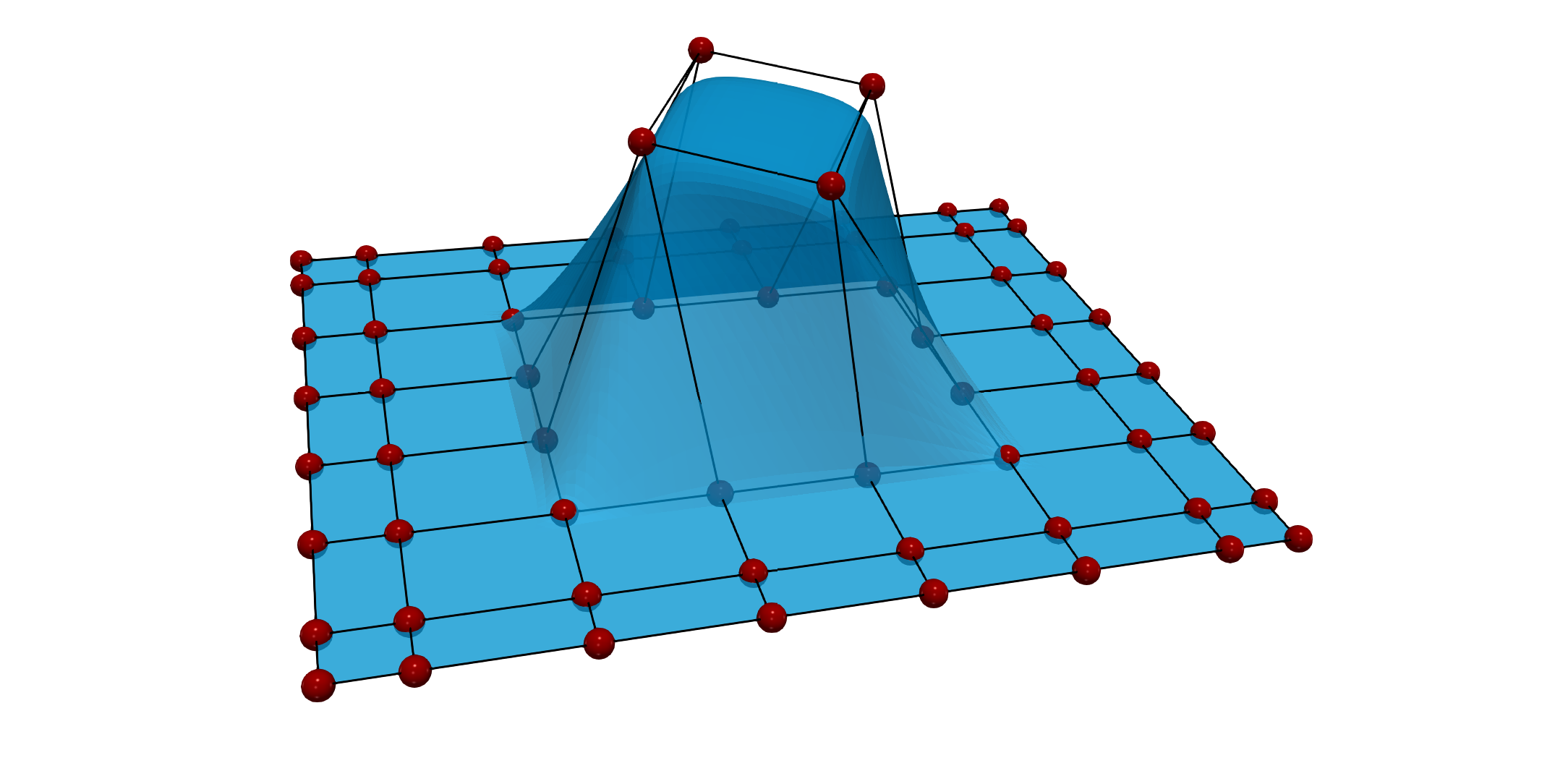}
	\caption{B-spline representation (blue) obtained from control points (red) for a bi-quadratic B-spline.
	The upper four control points are rotated, illustrating a possible deformation.}
	\label{fig:spline_example}
\end{figure}

Splines belong to the group of parametric shape representations.
Therefore, each coordinate in the parametric space is connected to one point in physical space.
This mapping is best understood using a simple \textit{B-spline} surface that is written as:
\begin{equation}
	\mathbf{S}\left(\xi,\eta\right) = \sum_{j=1}^{m}\sum_{i=1}^{n}N_{i,r}N_{j,p}\left(\xi\right)\mathbf{B}_{i,j},
	\label{eq:}
\end{equation}
where $\xi$ and $\eta$ denote the parametric coordinates (two for the surface), $N_{i,r}$ denote the interpolation or \textit{basis functions} of order $r$ in the first parametric direction, $N_{j,p}$ denote the basis functions of order $p$ in the second parametric direction, and finally, $\mathbf{B}$ denotes the support or \textit{control points}.
Figure~\ref{fig:spline_example} illustrates the concept and visualizes how single control points affect the geometry.
The control grid (i.e., the polygon spanned by the control point) aligns with the $\xi$ and $\eta$ directions, and any parametric coordinate (within the spline's parametric bounds) maps to one point of the blue shape.
Consequently, the spline mapping allows controlling an arbitrary number of parametric points by a constant, typically low, number of control points.
Being able to control a high number of points with few control points will be the basic idea of filtering using splines.\par

One can obtain geometry parameterizations from splines in multiple ways.
As shown in Fig.~\ref{fig:spline_example}, one way uses the B-splines as a boundary representation.
Such spline-based boundary representations are common in \gls{cad}.
Using these \gls{cad} representations, their control points (i.e., the red points in Fig.~\ref{fig:spline_example}) can be directly used as design variables in shape optimization.
However, this use of the CAD's geometry parameterization limits the design process because a given spline may not be able to represent shapes substantially different from the initial design.
Consequently, if modifications of the spline's parameterization, such as inserting additional control point lines, are to be avoided, this limitation restricts the use of the \gls{cad} spline to use cases that deal with small shape updates such as \textit{die} or \textit{mold design} \cite{Elgeti12}.\par

\begin{figure}[!h]
	\centering
	\tikzset{every picture/.style={line width=0.75pt}} 

	\begin{tikzpicture}[x=0.75pt,y=0.75pt,yscale=-1,xscale=1]

		\draw (75.5,144.72) node  {\includegraphics[width=52.5pt,height=97.83pt]{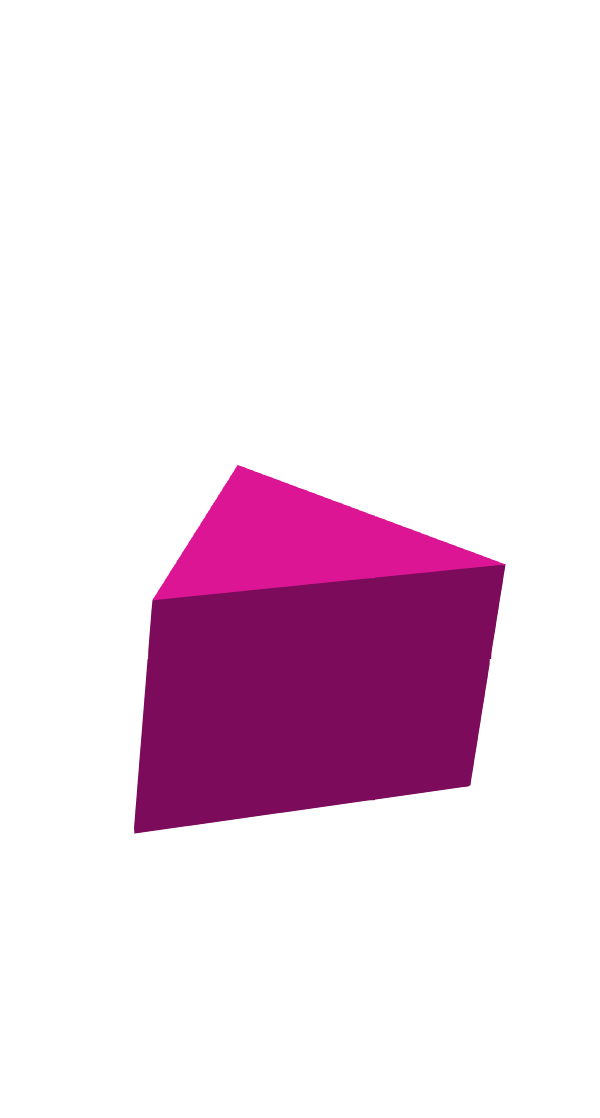}};
		\draw (175.66,144.78) node  {\includegraphics[width=70.06pt,height=97.91pt]{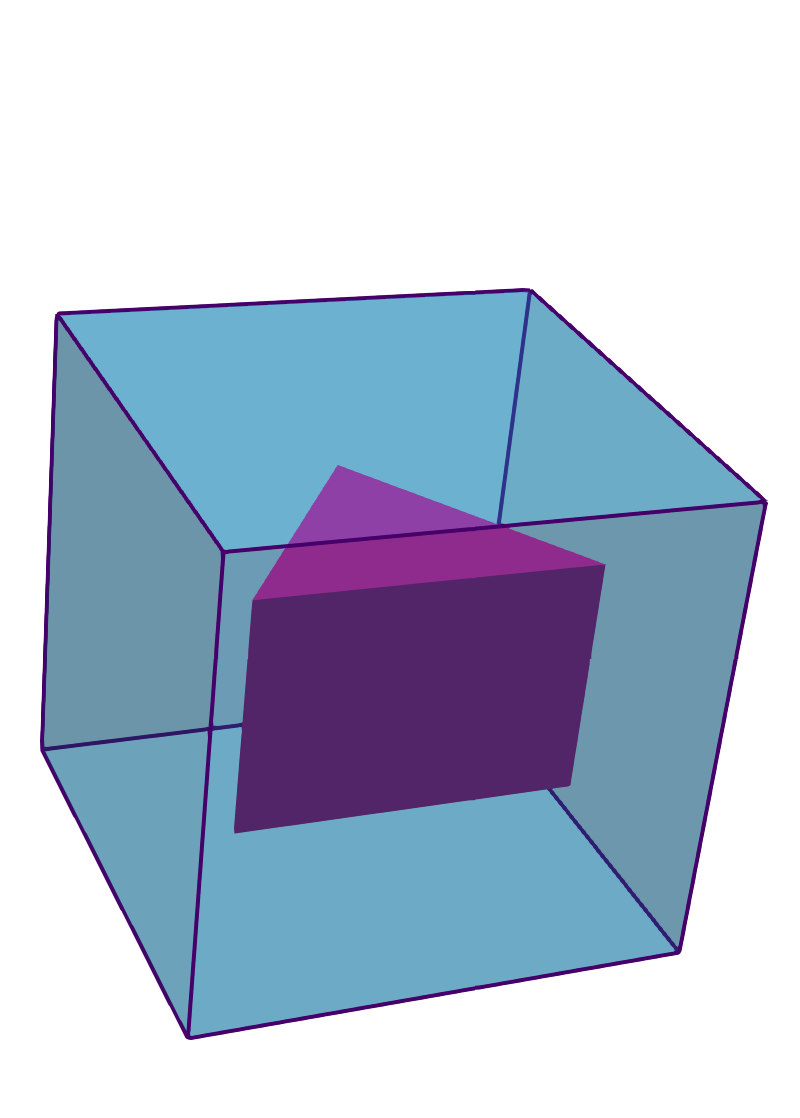}};
		\draw (296.53,144.78) node  {\includegraphics[width=70.06pt,height=97.91pt]{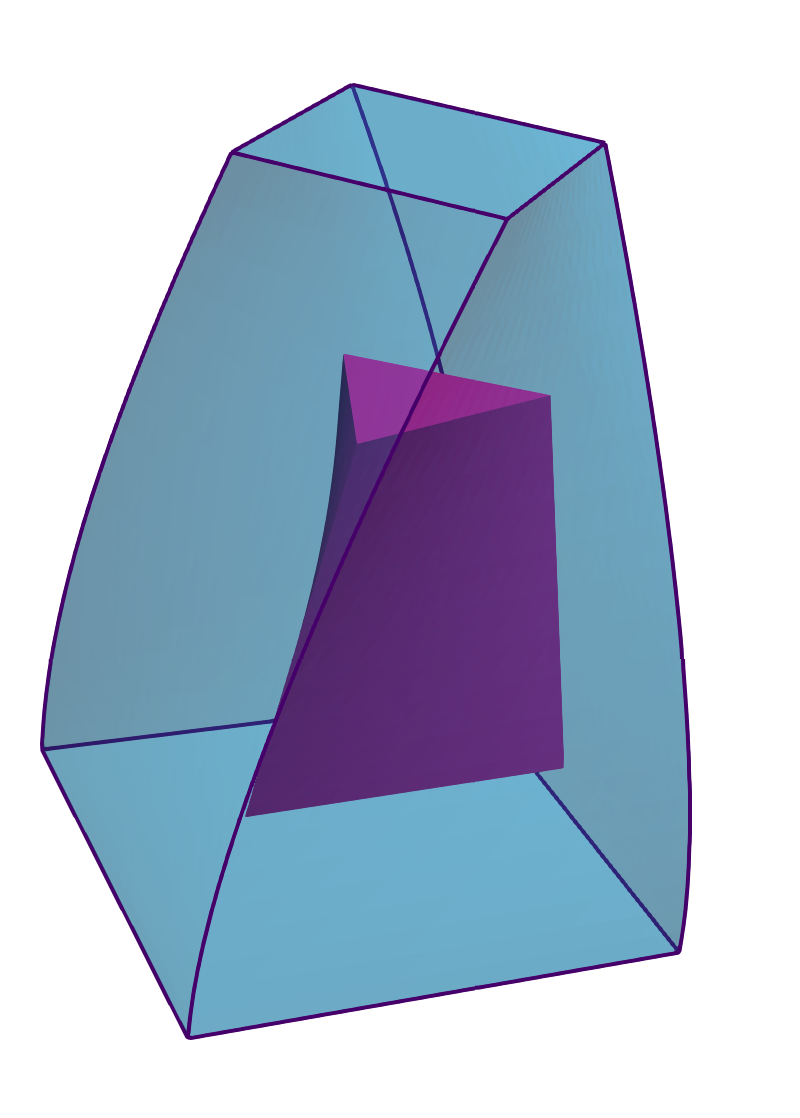}};
		\draw (387.68,144.72) node  {\includegraphics[width=52.5pt,height=97.83pt]{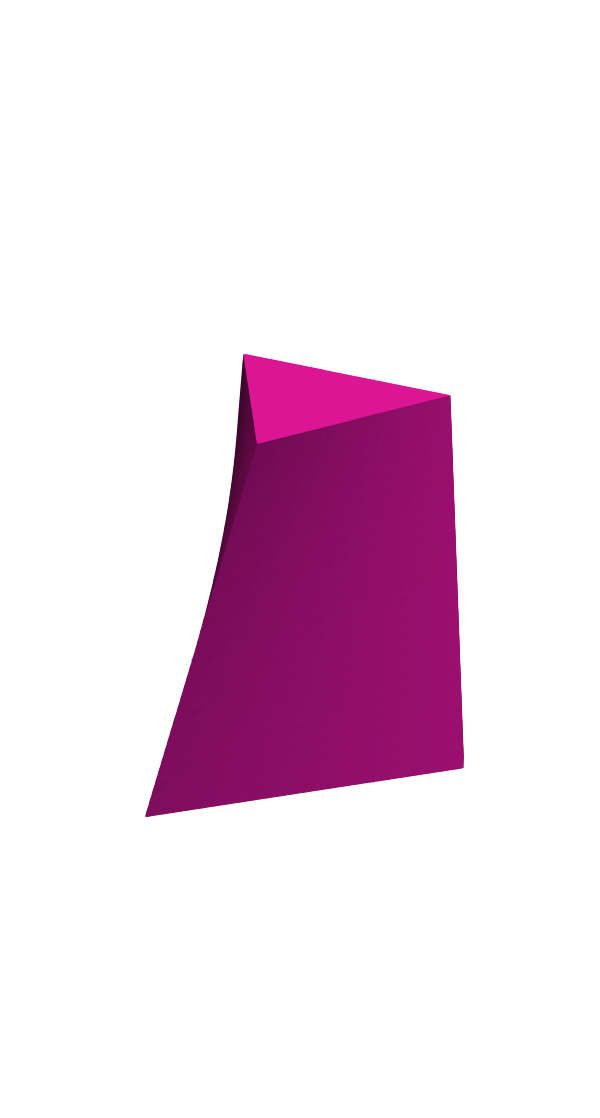}};
		\draw    (103.1,156.55) -- (122.1,156.55) ;
		\draw [shift={(125.1,156.55)}, rotate = 180] [fill={rgb, 255:red, 0; green, 0; blue, 0 }  ][line width=0.08]  [draw opacity=0] (5.36,-2.57) -- (0,0) -- (5.36,2.57) -- cycle    ;
		\draw    (223.6,156.55) -- (242.6,156.55) ;
		\draw [shift={(245.6,156.55)}, rotate = 180] [fill={rgb, 255:red, 0; green, 0; blue, 0 }  ][line width=0.08]  [draw opacity=0] (5.36,-2.57) -- (0,0) -- (5.36,2.57) -- cycle    ;
		\draw    (340.6,156.55) -- (359.6,156.55) ;
		\draw [shift={(362.6,156.55)}, rotate = 180] [fill={rgb, 255:red, 0; green, 0; blue, 0 }  ][line width=0.08]  [draw opacity=0] (5.36,-2.57) -- (0,0) -- (5.36,2.57) -- cycle  ;
	\end{tikzpicture}
	\caption{Free-Form Deformation using a volumetric spline (light blue) applied to a mixing element (pink). The control points are omitted in this figure.
		The embedded shape deforms correspondingly to the embedding, simple, volumetric spline.}
	\label{fig:FFD}
\end{figure}
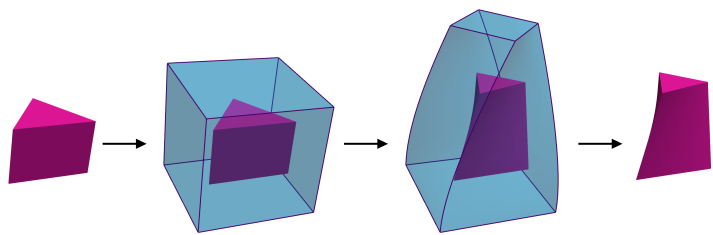

An alternative to using boundary B-splines is \gls{ffd} \cite{Sederberg1986}.
In \gls{ffd}, first, an -- often volumetric -- spline is constructed around the body to be deformed.
Second, this volumetric spline is deformed, and finally, the resulting deformation field is imposed on the enclosed body.
Fig.~\ref{fig:FFD} visualizes this process.
The advantage of \gls{ffd} is that the spline is constructed irrespective of the enclosed shape, which gives complete freedom in choosing degree and resolution.
This freedom allows tailoring the spline to the designer's needs (rather than using a given parameterization optimized for \gls{cad} usage).
Therefore, \gls{ffd} is widely applied, with just one example being the recent works by Lassila and Rozza combining \gls{ffd} and reduced order modeling \cite{RozzaFFD2010}.
A combination of both methods, boundary B-splines and \gls{ffd}, will be compared against the novel shape parameterization based on neural networks that use \gls{ffd} as a generic interface to modify any given \gls{cad} spline, which in turn is used to update the boundary of the simulation domain \cite{hube2021}.

\section{Shape parametrization using neural networks}
\label{sec:shppar}
As explained in Sec.~\ref{sec:geomfilters}, the prime objective of this work is to investigate how neural networks can be used to encode different shapes in a single set of a few continuous variables.
In order to train the network, thereby determining such a condensed representation, it has to be provided with suitable data.
\textit{Suitable} here means that the input data (i.e., shapes) are provided in such a way that the network can learn from this data.
In addition -- using the same data format -- we need to be able to produce high-quality computational meshes from the neural network's output.\par
In the following, we first introduce deep generative models and then describe a shape representation meeting these two requirements.
Finally, we discuss the training data generation and utilization of neural networks as shape generators.

\subsection{Deep generative models}
\label{subsec:deepGenMods}
With the advent of \textit{generative models}, an alternative approach to shape parameterization emerged.
In this subsection, we review two of the most common approaches of generative models, explain their basic concepts and use, and detail how they can be employed for geometric filtering.\par
Generative models are an application of neural networks and, thus, in essence, classification algorithms.
\textit{Classification} here means the ability to determine whether a certain object is in some measure \textit{close} to a specified input.
Conversely to just classifying input, such models can also be used to generate an output that resembles an input.
\textit{Resemble}, however, needs to be explained.
In most applications, the user is not interested in reproducing a given input exactly.
Instead, the output should only be \textit{like} the input (i.e., the output should feature a slight variation).
Generative models attempt to achieve this goal via statistical modeling.
An excellent guide to generative models is found in \cite{doersch2021tutorial}, with special focus on the \textit{Variational Autoencoder} (VAE).\par
The VAE, like the traditional autoencoder, consists of an \textit{encoder} and a \textit{decoder} and aims to reproduce any given data while passing the input through a bottleneck.
However, its probabilistic formulation using the so-called "reparametrization trick" provides an exceptional advantage over the traditional autoencoder in practice \cite{kingma2014autoencoding}.
The roles of the encoder and the decoder can be interpreted as two separate processes.
The encoder learns relations in the given data and encodes them in so-called \textit{latent variables}, $\mathbf{z}$.
Given these latent variables, the decoder, in turn, learns to produce data that is \textit{likely} to match the input.
Once trained, the user can omit the encoder and directly generate new data from sampling the latent space.
For details, we refer to \cite{ doersch2021tutorial, kingma2014autoencoding}, and for applications, we refer to \cite{VAE_shapeMorph, liu2018learning} and \cite{Tan_VAE4Shapes}. \par
\begin{figure}[h!]
	\centering
	\begin{subfigure}[b]{\textwidth}
		\centering

		\tikzset{every picture/.style={line width=0.75pt}} 

		\begin{tikzpicture}[x=0.75pt,y=0.75pt,yscale=-0.75,xscale=0.75]

			\draw  [line width=1.5]  (130.73,156.3) -- (209.73,193.3) -- (209.73,242.3) -- (130.73,279.3) -- cycle ;
			\draw  [fill={rgb, 255:red, 248; green, 231; blue, 28 }  ,fill opacity=1 ][line width=1.5]  (216.73,192.47) -- (231.92,192.47) -- (231.92,242.56) -- (216.73,242.56) -- cycle ;
			\draw [line width=0.75]    (80.98,216.47) -- (119.98,216.47) ;
			\draw [shift={(122.98,216.47)}, rotate = 180] [fill={rgb, 255:red, 0; green, 0; blue, 0 }  ][line width=0.08]  [draw opacity=0] (5.36,-2.57) -- (0,0) -- (5.36,2.57) -- cycle    ;
			\draw  [line width=1.5]  (317.92,279.3) -- (238.92,243.06) -- (238.92,192.54) -- (317.92,156.3) -- cycle ;
			\draw (43.8,216.8) node  {\includegraphics[width=40pt,height=40pt]{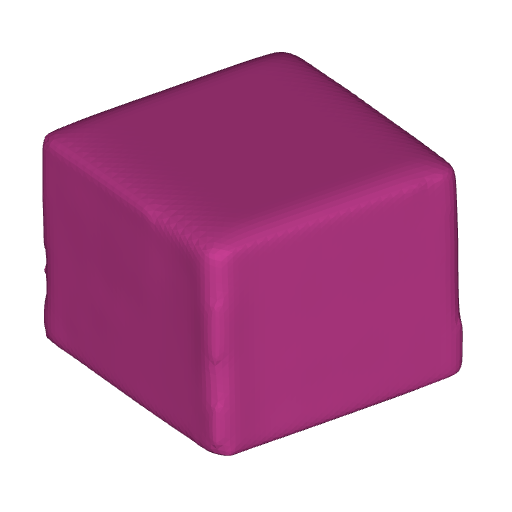}};
			\draw (402.8,215.8) node  {\includegraphics[width=40pt,height=40pt]{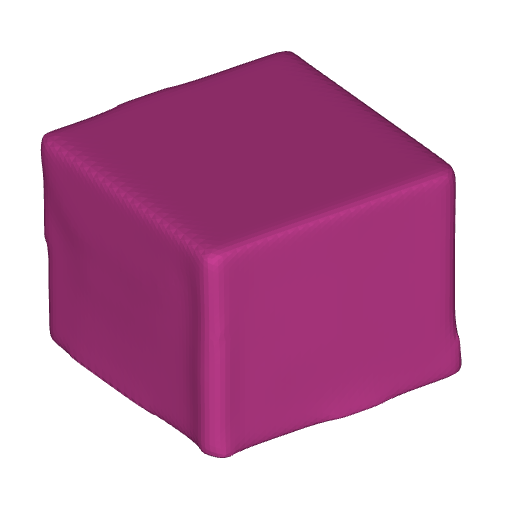}};
			\draw [line width=0.75]    (326.98,216.47) -- (365.98,216.47) ;
			\draw [shift={(368.98,216.47)}, rotate = 180] [fill={rgb, 255:red, 0; green, 0; blue, 0 }  ][line width=0.08]  [draw opacity=0] (5.36,-2.57) -- (0,0) -- (5.36,2.57) -- cycle    ;

			\draw (170.95,215.5) node   [align=left] {Encoder};
			\draw (279.37,215.5) node   [align=left] {Decoder};
		\end{tikzpicture}
		\caption{Autoencoder providing an input-to-output mapping while passing data through a bottleneck, i.e., the low-dimensional latent representation.}
		\label{fig:autoencoder}
	\end{subfigure}
	\begin{subfigure}[b]{\textwidth}
		\centering

		\tikzset{every picture/.style={line width=0.75pt}} 

		\begin{tikzpicture}[x=0.75pt,y=0.75pt,yscale=-0.8,xscale=0.8]

			\draw  [fill={rgb, 255:red, 248; green, 231; blue, 28 }  ,fill opacity=1 ][line width=1.5]  (18.55,97.47) -- (33.73,97.47) -- (33.73,147.56) -- (18.55,147.56) -- cycle ;
			\draw  [line width=1.5]  (18.2,201.19) -- (119.2,201.19) -- (119.2,262.12) -- (18.2,262.12) -- cycle ;
			\draw [line width=0.75]    (374.98,170.47) -- (389.9,170.47) ;
			\draw [shift={(392.9,170.47)}, rotate = 180] [fill={rgb, 255:red, 0; green, 0; blue, 0 }  ][line width=0.08]  [draw opacity=0] (5.36,-2.57) -- (0,0) -- (5.36,2.57) -- cycle    ;
			\draw [line width=0.75]    (221.73,118.67) .. controls (244.44,119.12) and (216.48,167.67) .. (254.87,170.55) ;
			\draw [shift={(257.3,170.67)}, rotate = 181.36] [fill={rgb, 255:red, 0; green, 0; blue, 0 }  ][line width=0.08]  [draw opacity=0] (5.36,-2.57) -- (0,0) -- (5.36,2.57) -- cycle    ;
			\draw [line width=0.75]    (221.73,230.47) .. controls (244.44,230.92) and (216.48,171.9) .. (254.87,170.48) ;
			\draw [shift={(257.3,170.47)}, rotate = 181.36] [fill={rgb, 255:red, 0; green, 0; blue, 0 }  ][line width=0.08]  [draw opacity=0] (5.36,-2.57) -- (0,0) -- (5.36,2.57) -- cycle    ;
			\draw   (396.4,145.97) .. controls (396.4,143.56) and (398.36,141.6) .. (400.77,141.6) -- (413.89,141.6) .. controls (416.3,141.6) and (418.26,143.56) .. (418.26,145.97) -- (418.26,195.95) .. controls (418.26,198.36) and (416.3,200.32) .. (413.89,200.32) -- (400.77,200.32) .. controls (398.36,200.32) and (396.4,198.36) .. (396.4,195.95) -- cycle ;
			\draw  [fill={rgb, 255:red, 208; green, 2; blue, 27 }  ,fill opacity=1 ] (400.33,158.56) .. controls (400.33,154.72) and (403.45,151.6) .. (407.3,151.6) .. controls (411.14,151.6) and (414.26,154.72) .. (414.26,158.56) .. controls (414.26,162.41) and (411.14,165.53) .. (407.3,165.53) .. controls (403.45,165.53) and (400.33,162.41) .. (400.33,158.56) -- cycle ;
			\draw  [line width=1.5]  (117.92,184) -- (38.92,147.76) -- (38.92,97.24) -- (117.92,61) -- cycle ;
			\draw (184,229.8) node  {\includegraphics[width=40pt,height=40pt]{figs/interpolation/4/image1.png}};
			\draw (182,122) node  {\includegraphics[width=40pt,height=40pt]{figs/interpolation/16/image1.png}};
			\draw  [line width=1.5]  (262.98,108.86) -- (368.55,145.16) -- (368.55,195.56) -- (262.98,231.86) -- cycle ;
			\draw  [fill={rgb, 255:red, 126; green, 211; blue, 33 }  ,fill opacity=1 ] (400.73,183.36) .. controls (400.73,179.52) and (403.85,176.4) .. (407.7,176.4) .. controls (411.54,176.4) and (414.66,179.52) .. (414.66,183.36) .. controls (414.66,187.21) and (411.54,190.33) .. (407.7,190.33) .. controls (403.85,190.33) and (400.73,187.21) .. (400.73,183.36) -- cycle ;
			\draw [line width=0.75]    (129.41,118.67) -- (144.2,118.67) ;
			\draw [shift={(147.2,118.67)}, rotate = 180] [fill={rgb, 255:red, 0; green, 0; blue, 0 }  ][line width=0.08]  [draw opacity=0] (5.36,-2.57) -- (0,0) -- (5.36,2.57) -- cycle    ;
			\draw [line width=0.75]    (128.98,229.67) -- (143.78,229.67) ;
			\draw [shift={(146.78,229.67)}, rotate = 180] [fill={rgb, 255:red, 0; green, 0; blue, 0 }  ][line width=0.08]  [draw opacity=0] (5.36,-2.57) -- (0,0) -- (5.36,2.57) -- cycle    ;

			\draw (78.79,120.5) node   [align=left] {Generator};
			\draw (68.7,231.66) node   [align=left] {Real world\\data};
			\draw (315.77,170.36) node   [align=left] {Discriminator};
			\draw (482.16,158.5) node   [align=left] {Generated data};
			\draw (481.97,182.1) node   [align=left] {Real world data};

		\end{tikzpicture}

		\caption{Generative adversarial model learning latent space by inferring representations that enable generating output indistinguishable from the input.}
		\label{fig:gan}
	\end{subfigure}
    \caption{Two main concepts of deep generative networks: Variational Autoencoders and Generative Adversarial Networks.}
\end{figure}

The difference between the spline-based approach and generative models is the choice of latent variables.
When the human designer creates a spline parameterization that allows modifying geometry in the desired way, the optimization variables are the control points, which are \textit{intuitively} placed in $\mathbb{R}^3$ by the designer.
Generative models, in contrast, \textit{learn} a latent space and explicitly assume that the single latent variables do not have an intuitive interpretation.
As a result, data is compressed from a high dimensional intuitive design space, in our case $\chi \subset \mathbb{R}^{3\times n}$, onto a hardly interpretable, feature-dense, low-dimensional latent space $Z$.
In short, generative models use the computational power of neural networks to find a dense classification space that one can sample to produce new data.
For the VAE, this process is depicted in Fig. \ref{fig:autoencoder}.\par

A competing concept to VAEs are \textit{Generative Adversarial Networks (GANs)}.
Their basic structure is shown in Fig.~\ref{fig:gan}.
GANs, first introduced by Goodfellow \textit{et al.} \cite{Goodfellow2014}, follow a different concept and train two adversarial nets, the \textit{generator} and the \textit{discriminator}.
In GANs, the generator is trained to create data that mimics real-world data, while the discriminator tries to determine whether or not a dataset was artificially created.
In a minimax fashion, the generator's learning goal is to maximize the probability of the discriminator making a wrong decision.\par
GANs have proven to be an excellent tool for shape modeling.
Wu \textit{et al.}, for example, apply a GAN for 3D shape generation and demonstrate their superior performance compared to three-dimensional VAEs.
They even use a GAN to reconstruct three-dimensional models from two-dimensional images based on the a VAE output that is used to infer a latent representation for these images\cite{Wu2016}.
As in \cite{liu2018learning}, Wu \textit{et al.} also demonstrate the ability to apply shape interpolation and shape arithmetic to the learned latent representation.
More recently, Ramasinghe \textit{et al.} \cite{ramasinghe2020} utilize a GAN to model high-resolution three-dimensional shapes using point clouds.

\subsection{Implicit shape representation}
The neural network learns a mapping between the low-dimensional latent space and a three-dimensional body.
In order to construct such a mapping, we first need to define how to represent our shapes (i.e., define what data the neural network actually has to learn).
Before presenting the approach chosen in this work, we review standard methods and their limitation.\par
Three ways of shape representation are common in machine learning: (1) voxels, (2) point clouds, and (3) meshes \cite{Park_deepsdf}.
The problem with meshes is that the mesh topology also prescribes the possible shape topologies.
Point clouds, in contrast, can represent arbitrary topologies, but prescribe a given resolution.
Finally, voxels can represent arbitrary topologies and vary in resolution, but, unfortunately, the memory consumption scales cubically with the resolution.
Because of these drawbacks, the network utilized in this work learns \glspl{sdf} following a network configuration originally proposed by Park \textit{et al.} \cite{Park_deepsdf}. \par
\Glspl{sdf} provide the distance to the closest point on the to-be-encoded surface for every point in space.
Furthermore, encoded in the sign, information on whether the point lies inside or outside the surface is available.
Using such continuous \gls{sdf} data, a shape is then extracted -- at an arbitrary resolution suitable for meshing -- as its zero-valued isosurface.

\subsection{Training set generation}
\label{sec:trainig_set_generation}
As mentioned in Sec.~\ref{subsec:deepGenMods}, training a neural network requires a set of source shapes.
However, to the authors' knowledge, no shape library exists for mixing elements in single screw extruders.
Thus, we explain an approach to building custom training sets.\par 
To generate a suitable training set, we first select the basis shapes that should be considered -- pin and pineapple mixers in our case.
From this choice, we arbitrarily infer a total of four basis shapes (i.e., triangle, square, hexagon and cylinder -- cf. Fig. \ref{fig:shape_variations}) -- clearly too few for successful and meaningful training.
The basis shapes are varied using (a combination of) \glspl{ffd} to gather an appropriate number of training shapes.
Examples of applied deformations are given in Fig.~\ref{fig:shape_variations}.
In total, 2659 training shapes are generated.
\def  \subfigwidth {0.24\linewidth}
\begin{figure}[!h]
	\centering
		\lineskip=0pt
	\begin{subfigure}{0.875\linewidth}
	\lineskip=0pt

	\begin{subfigure}[t]{\subfigwidth}
		\centering
		\includegraphics[width=0.45\linewidth]{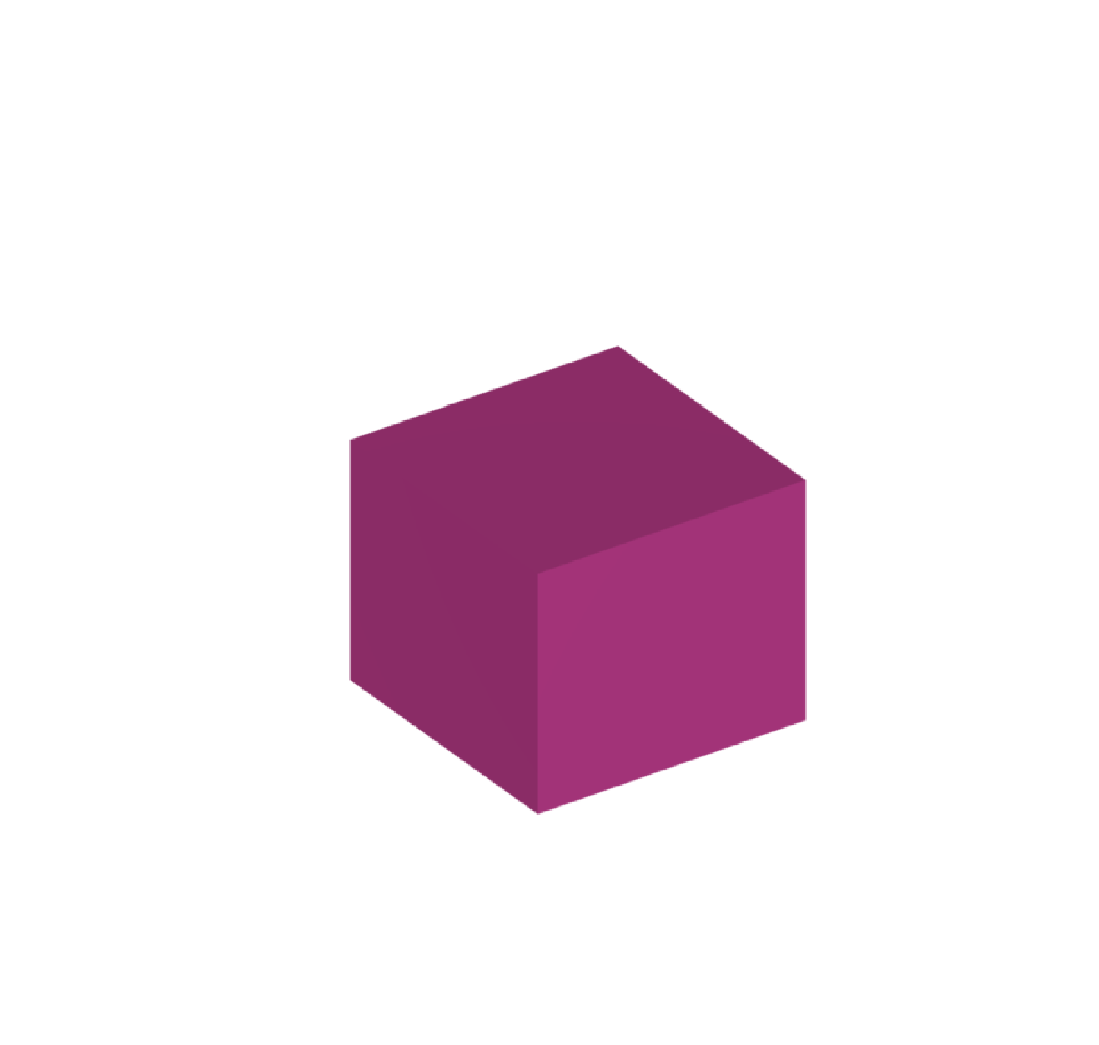}
		\caption{Square base}
	\end{subfigure}
	\hfill
	\begin{subfigure}[t]{\subfigwidth}
		\centering
		\includegraphics[width=0.45\linewidth]{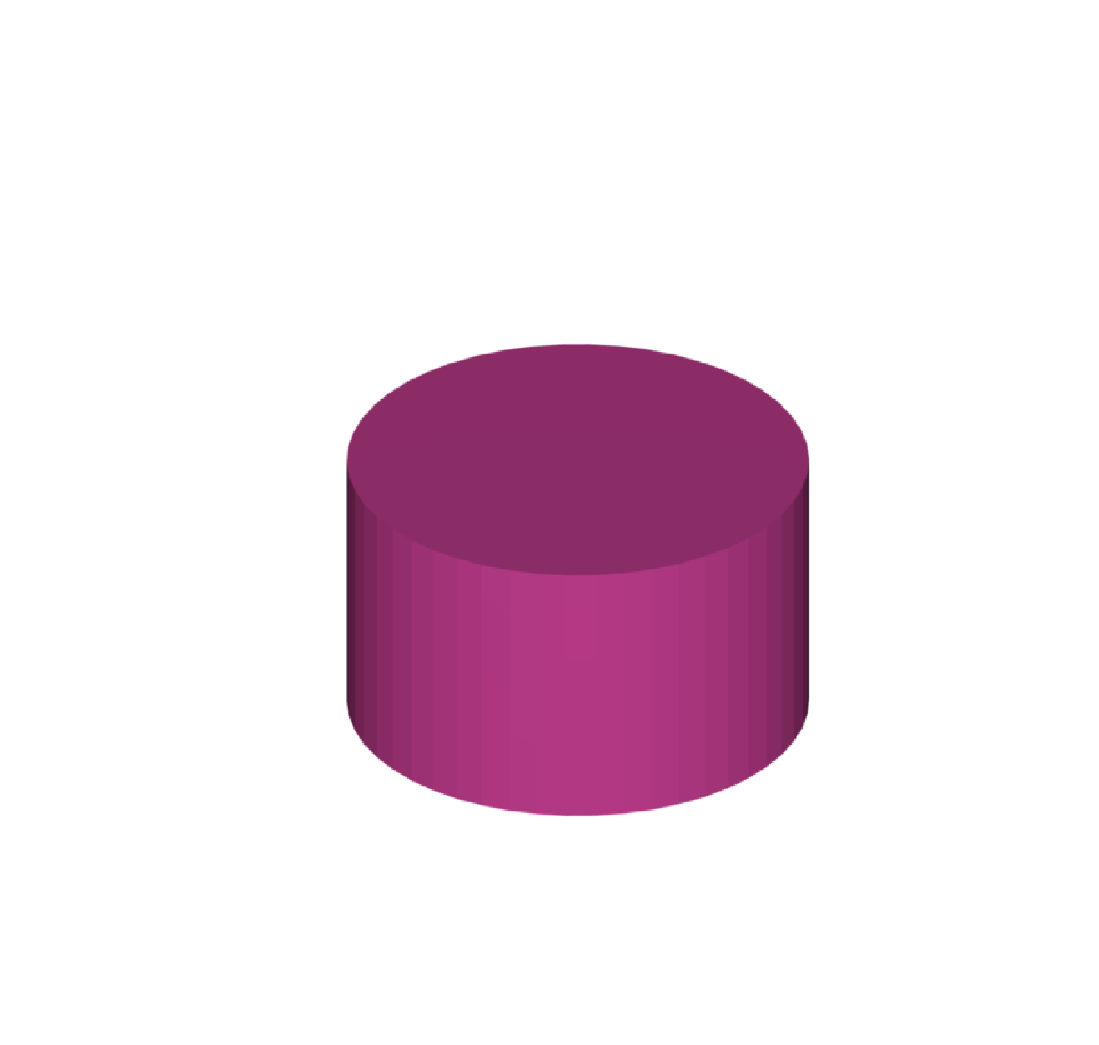}
		\caption{Cylinder base}
	\end{subfigure}
	\hfill
	\begin{subfigure}[t]{\subfigwidth}
		\centering
		\includegraphics[width=0.45\linewidth]{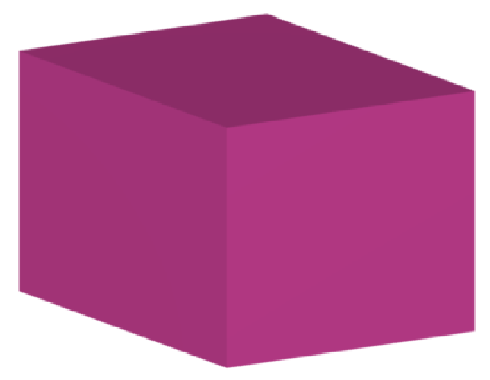}
		\caption{Shrink along $x$}
	\end{subfigure}
	\hfill
	\begin{subfigure}[t]{\subfigwidth}
		\centering
		\includegraphics[width=0.45\linewidth]{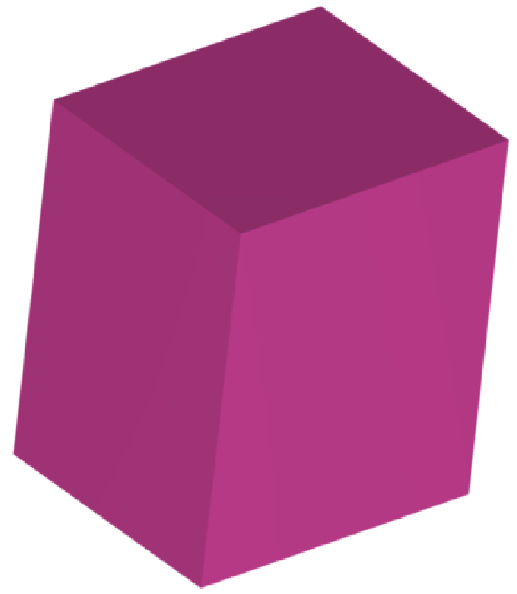}
		\caption{Translate top $x$}
	\end{subfigure}
	\\
	\begin{subfigure}[t]{\subfigwidth}
		\centering
		\includegraphics[width=0.45\linewidth]{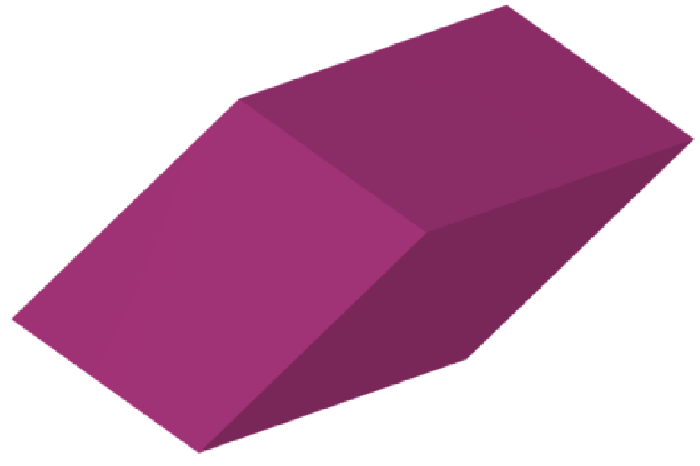}
		\caption{Tranlate top $y$}
	\end{subfigure}
	\hfill
	\begin{subfigure}[t]{\subfigwidth}
		\centering
		\includegraphics[width=0.45\linewidth]{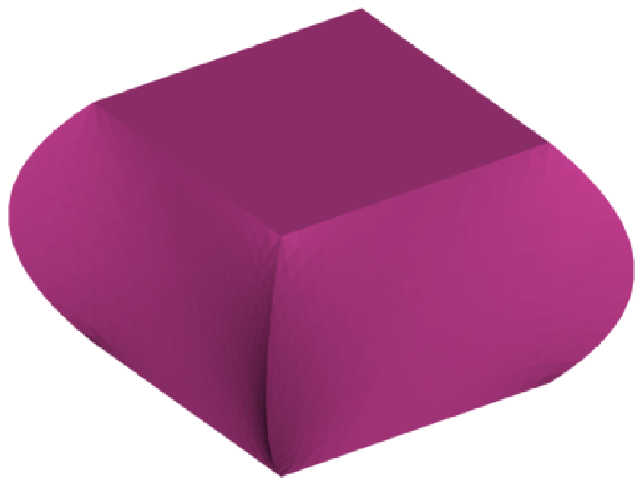}
		\caption{Expand middle}
	\end{subfigure}
	\hfill
	\begin{subfigure}[t]{\subfigwidth}
		\centering
		\includegraphics[width=0.45\linewidth]{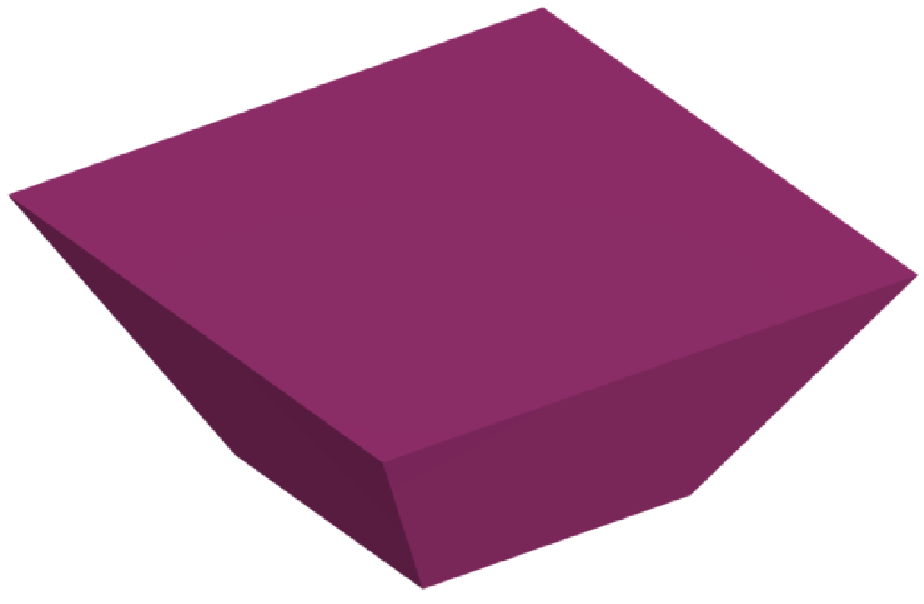}
		\caption{Expand top}
	\end{subfigure}
	\hfill
	\begin{subfigure}[t]{\subfigwidth}
		\centering
		\includegraphics[width=0.45\linewidth]{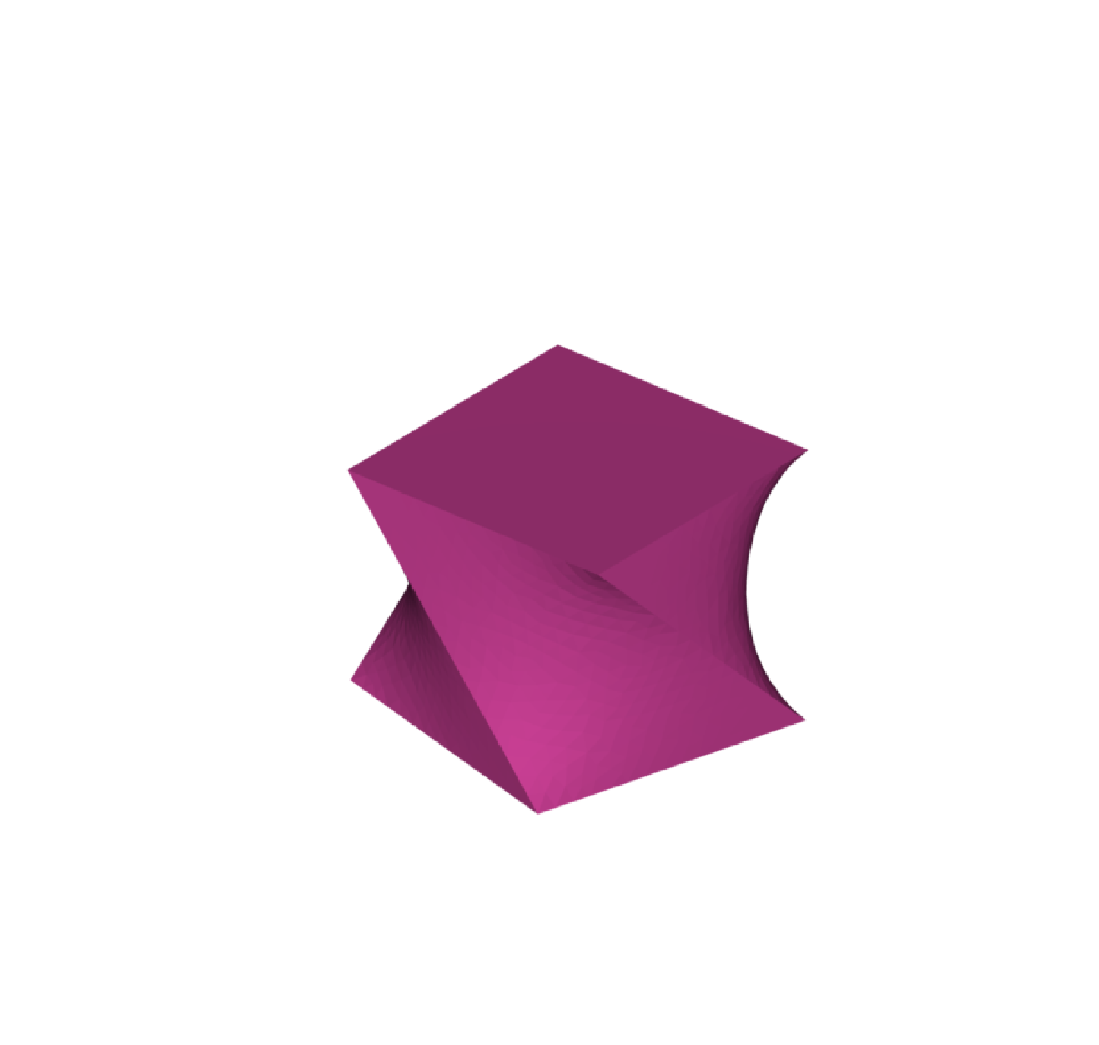}
		\caption{Twist top}
	\end{subfigure}
	\\

	\end{subfigure}
	\caption{Examples of basic shapes and applied deformations. In total, a triangle, a square, a cylinder, and a hexagon are used as basis shapes.}
	\label{fig:shape_variations}
\end{figure}

To obtain \gls{sdf}-training data from these shapes, we follow the approach by Park \textit{et al.} \cite{Park_deepsdf} and first normalize each shape to fit into a unit sphere.
Then, we sample 500,000 spatial points and their \gls{sdf} value pairs using the trimesh library \cite{trimesh}.


\subsection{Shape generator}
As explained, the shape generator's task is to provide a mixing element given a set of optimization variables.
The shape generator -- in this work -- is thus built around the neural network, which is presented in the following.\par
The utilized neural network is based on DeepSDF auto-decoder \cite{Park_deepsdf}: a feed-forward network with ten fully connected layers, with each of the eight hidden (i.e., internal) layers having 256 neurons and ReLU activation functions.
In contrast to auto-encoders, the auto-decoder only trains the decoder using a simultaneous optimization of the network parameters and the latent code during training.
We investigate four, eight, and sixteen as latent dimensions, $l$.
The input layer consists of these $l$ neurons concatenated with a three-dimensional query location.
The output layer has only one neuron with a $\tanh$ activation function.
For details on the chosen \gls{sdf} network, we, again, refer to \cite{Park_deepsdf}.
To train the network, we use the ADAM optimization algorithm \cite{kingma2017adam}.
To utilize improved learning rates, we follow a progressive approach with initial rates $\varepsilon_0=5e-4$ for $\vekt{\theta}$, and $\varepsilon_0=1e-3$ for $\vekt{z}$, and a decay as:
\begin{equation}
	\varepsilon = \varepsilon_0\cdot\left(0.5^{e\%500}\right),
\end{equation}
where $e$ denotes the current training iteration (i.e., \textit{epochs}) -- and $\%$ denotes integer division.
The network's training can be seen as the parametrization of the shapes.\par
To extract isosurfaces (i.e., to generate new mixing elements) from the trained network's \gls{sdf} output, we sample a discrete \gls{sdf} field and apply a marching cube algorithm \cite{Lorensen_mc} in the implementation of \cite{Lewiner_efficient_mc}.
Finally, we apply automated meshing using TetGen \cite{si_tetgen} to obtain a simulation domain as depicted in Fig.~\ref{fig:sim_domain}, including the new mixing element.

\section{The developed shape optimization framework}
\label{sec:compSetup}
In general, our framework consists of three building blocks: (1) \textit{shape generator}, (2) \textit{flow solver}, and (3) \textit{optimizer}, which will be described in the following.\par
Starting with an initial set of optimization variables, $\boldsymbol{\sigma}_0$, the shape generator creates a new mixing element $\Omega\left(\boldsymbol{\sigma}_0\right)$.
The flow solver then computes the flow field around this mixing element, which the optimizer evaluates to determine the flow's degree of mixing.
Based on the obtained mixing value and by comparison to previous iterations, an optimization algorithm determines a new set of optimization variables.
This sequence is iteratively re-run until either a maximum number of iterations is reached or any other termination criterion -- typically a good objective value or insignificant objective decrease -- is met.

\subsection{Flow solver and simulation model}
\label{subsec:simulation_model}
The flow solver and simulation model is identical to the one introduced in \cite{hube2021} and therefore only summarized in the following. 
The flow field induced by the various mixing elements is obtained from solving the steady, incompressible non-isothermal Navier-Stokes equations using a Carreau model and WLF temperature correction.
The governing equations are discretized with linear stabilized finite elements and solved using a Newton linearization and a GMRES iterative solver.
Subsequently, we solve a set of advection equations using the identical configuration to mimic particle tracking, which we use as an input to our objective function.
All methods are implemented in an in-house flow solver.\par
\begin{figure}[!h]
	\centering
	\tikzset{every picture/.style={line width=0.75pt}} 

	\begin{tikzpicture}[x=0.75pt,y=0.75pt,yscale=-0.75,xscale=0.75]

		\draw [color={rgb, 255:red, 74; green, 144; blue, 226 }  ,draw opacity=1 ]   (300.27,148.54) -- (272.27,201.54) ;
		\draw [color={rgb, 255:red, 74; green, 144; blue, 226 }  ,draw opacity=1 ]   (272.4,190.45) -- (277.34,188.61) ;
		\draw [shift={(280.15,187.55)}, rotate = 159.48] [fill={rgb, 255:red, 74; green, 144; blue, 226 }  ,fill opacity=1 ][line width=0.08]  [draw opacity=0] (5.36,-2.57) -- (0,0) -- (5.36,2.57) -- cycle    ;
		\draw [color={rgb, 255:red, 74; green, 144; blue, 226 }  ,draw opacity=1 ]   (272.8,182.45) -- (281.71,179.34) ;
		\draw [shift={(284.55,178.35)}, rotate = 160.76] [fill={rgb, 255:red, 74; green, 144; blue, 226 }  ,fill opacity=1 ][line width=0.08]  [draw opacity=0] (5.36,-2.57) -- (0,0) -- (5.36,2.57) -- cycle    ;
		\draw [color={rgb, 255:red, 74; green, 144; blue, 226 }  ,draw opacity=1 ]   (272.4,174.85) -- (286.11,170.13) ;
		\draw [shift={(288.95,169.15)}, rotate = 160.99] [fill={rgb, 255:red, 74; green, 144; blue, 226 }  ,fill opacity=1 ][line width=0.08]  [draw opacity=0] (5.36,-2.57) -- (0,0) -- (5.36,2.57) -- cycle    ;
		\draw [color={rgb, 255:red, 74; green, 144; blue, 226 }  ,draw opacity=1 ]   (272.4,166.85) -- (291.32,160.15) ;
		\draw [shift={(294.15,159.15)}, rotate = 160.5] [fill={rgb, 255:red, 74; green, 144; blue, 226 }  ,fill opacity=1 ][line width=0.08]  [draw opacity=0] (5.36,-2.57) -- (0,0) -- (5.36,2.57) -- cycle    ;

		\draw  [fill={rgb, 255:red, 171; green, 171; blue, 171 }  ,fill opacity=1 ] (286.77,115.29) -- (286.77,152.96) -- (258.02,163.71) -- (258.02,126.04) -- cycle ;
		\draw  [fill={rgb, 255:red, 125; green, 125; blue, 125 }  ,fill opacity=1 ] (258.02,126.04) -- (258.02,163.71) -- (218.02,148.79) -- (218.52,111.29) -- cycle ;
		\draw  [fill={rgb, 255:red, 228; green, 228; blue, 228 }  ,fill opacity=1 ] (286.77,115.29) -- (258.02,126.04) -- (218.52,111.29) -- (247.27,100.54) -- cycle ;
		\draw   (272.39,158.33) -- (272.27,201.54) -- (142.02,133.79) -- (141.77,89.79) -- cycle ;
		\draw   (390.39,115.58) -- (390.27,158.79) -- (260.02,91.04) -- (259.77,47.04) -- cycle ;
		\draw   (390.39,115.58) -- (390.27,158.79) -- (272.27,201.54) -- (272.39,158.33) -- cycle ;
		\draw   (259.77,47.04) -- (260.02,91.04) -- (142.02,133.79) -- (141.77,89.79) -- cycle ;
		\draw    (269.87,202.74) -- (250.01,209.85) ;
		\draw    (139.87,134.74) -- (120.01,141.85) ;
		\draw    (139.87,90.74) -- (120.01,97.85) ;
		\draw    (287.61,209.45) -- (274.41,202.65) ;
		\draw    (405.21,166.65) -- (392.01,159.85) ;
		\draw    (125.39,141.56) -- (252.64,207.33) ;
		\draw [shift={(254.41,208.25)}, rotate = 207.33] [fill={rgb, 255:red, 0; green, 0; blue, 0 }  ][line width=0.08]  [draw opacity=0] (12,-3) -- (0,0) -- (12,3) -- cycle    ;
		\draw [shift={(123.61,140.65)}, rotate = 27.33] [fill={rgb, 255:red, 0; green, 0; blue, 0 }  ][line width=0.08]  [draw opacity=0] (12,-3) -- (0,0) -- (12,3) -- cycle    ;
		\draw    (123.61,138.65) -- (123.61,99.05) ;
		\draw [shift={(123.61,97.05)}, rotate = 90] [fill={rgb, 255:red, 0; green, 0; blue, 0 }  ][line width=0.08]  [draw opacity=0] (12,-3) -- (0,0) -- (12,3) -- cycle    ;
		\draw [shift={(123.61,140.65)}, rotate = 270] [fill={rgb, 255:red, 0; green, 0; blue, 0 }  ][line width=0.08]  [draw opacity=0] (12,-3) -- (0,0) -- (12,3) -- cycle    ;
		\draw    (285.49,206.33) -- (398.54,164.93) ;
		\draw [shift={(400.41,164.25)}, rotate = 159.89] [fill={rgb, 255:red, 0; green, 0; blue, 0 }  ][line width=0.08]  [draw opacity=0] (12,-3) -- (0,0) -- (12,3) -- cycle    ;
		\draw [shift={(283.61,207.02)}, rotate = 339.89] [fill={rgb, 255:red, 0; green, 0; blue, 0 }  ][line width=0.08]  [draw opacity=0] (12,-3) -- (0,0) -- (12,3) -- cycle    ;

		\draw (50,108) node [anchor=north west][inner sep=0.75pt]   [align=left] {\SI{0.0075}{\meter}};
		\draw (133,179) node [anchor=north west][inner sep=0.75pt]   [align=left] {\SI{0.0315}{\meter}};
		\draw (344.01,188.63) node [anchor=north west][inner sep=0.75pt]   [align=left] {\SI{0.02405}{\meter}};

	\end{tikzpicture}

	\caption{Simulation domain with single mixing element resembling the flow around a single mixing element in the unwound screw channel. Flow conditions are shown in blue using a barrel rotation setup. For a detailed description of the objective function and governing equations, we refer the reader to \cite{hube2021}.
	}
	\label{fig:sim_domain}
\end{figure}
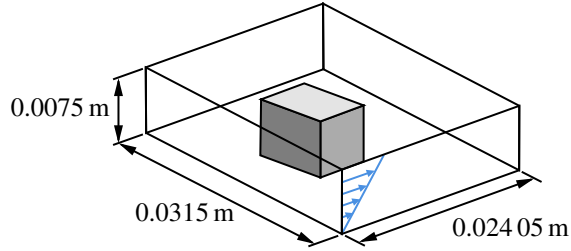

We make two simplifications to our simulation model (i.e., the single-screw-extruder flow channel): First, we simulate the flow around only a single mixing element instead of simulating the entire mixing section.
Second, we assume barrel rotation in an unwound flow channel section.
Both assumptions yield significantly reduced computational costs while allowing a qualitative mixing improvement.
To assess mixing, we mimic particle tracking by solving a series of advection equations yielding an inflow-outflow mapping for particles advected by the melt flow.
We process this advection information by subdividing a portion of the inflow domain into smaller rectangular subdomains.
In each of these rectangles, we select a set of particles such that the particle set's bounding box coincides with the rectangular subdomain.
Then, we follow each particle as they are conveyed through the domain, store each particle's position at the outflow domain, and finally construct a convex hull at the outflow around the same sets of points.
Averaging the convex hull's length increments between in- and outflow yields a simple yet robust objective function inspired by interfacial area measurements.
Using this objective function, we found that such a simulation model provides a good balance between accuracy and computational efficiency \cite{hube2021}.
Fig.~\ref{fig:sim_domain} depicts the chosen simulation domain.

\subsection{Optimizer}
We utilize the open-source optimization library Dakota \cite{Dakota} to drive the design process.
Two different algorithms are selected and described in the following.
The first algorithm is the Dividing RECTangle (DIRECT) algorithm, first introduced in \cite{Jones_direct}.
DIRECT belongs to the category of \textit{branch-and-bound} methods and uses n-dimensional trisection to iteratively partition the design space.
To find minima, it follows the approach of Lipschitzian optimization, which identifies the design space partition that should be further sampled by evaluating a lower bound to the objective value in each partition.
The partition with the lowest lower bound is chosen and further sampled.
DIRECT modifies that concept and computes multiple lower bounds that weight the current sampling value (i.e., the objective value in the partition center).
This promotes to further sample partitions with good objective values against the partition size, which permits to effectively sample large areas of unexplored design space. 
Thereby, DIRECT identifies \textit{multiple} partitions that are \textit{possibly optimal} and allows for global convergence.\par
\begin{figure}[!h]
	\centering
	\tikzset{every picture/.style={line width=0.75pt}} 

	\begin{tikzpicture}[x=0.75pt,y=0.75pt,yscale=-1,xscale=1]

		\draw  [fill={rgb, 255:red, 74; green, 144; blue, 226 }  ,fill opacity=0.14 ] (28.75,28.6) -- (450,28.6) -- (450,100.25) -- (28.75,100.25) -- cycle ;
		\draw   (30,120.5) -- (450,120.5) -- (450,260) -- (30,260) -- cycle ;
		\draw  [fill={rgb, 255:red, 255; green, 255; blue, 255 }  ,fill opacity=1 ] (36.7,38.25) -- (106.25,38.25) -- (106.25,60.03) -- (36.7,60.03) -- cycle ;
		\draw  [fill={rgb, 255:red, 255; green, 255; blue, 255 }  ,fill opacity=1 ] (37.25,70.03) -- (106.8,70.03) -- (106.8,91.78) -- (37.25,91.78) -- cycle ;
		\draw    (70.07,60.25) -- (70.07,66.5) ;
		\draw [shift={(70.07,69.5)}, rotate = 270] [fill={rgb, 255:red, 0; green, 0; blue, 0 }  ][line width=0.08]  [draw opacity=0] (5.36,-2.57) -- (0,0) -- (5.36,2.57) -- cycle    ;
		\draw    (70.39,100.25) -- (70.48,117.5) ;
		\draw [shift={(70.5,120.5)}, rotate = 269.7] [fill={rgb, 255:red, 0; green, 0; blue, 0 }  ][line width=0.08]  [draw opacity=0] (6.25,-3) -- (0,0) -- (6.25,3) -- cycle    ;
		\draw  [fill={rgb, 255:red, 74; green, 144; blue, 226 }  ,fill opacity=0.14 ] (37.25,130.03) -- (118.57,130.03) -- (118.57,151.78) -- (37.25,151.78) -- cycle ;
		\draw   (37.54,161.53) -- (118.86,161.53) -- (118.86,183.28) -- (37.54,183.28) -- cycle ;
		\draw    (75.63,152.25) -- (75.63,158.5) ;
		\draw [shift={(75.63,161.5)}, rotate = 270] [fill={rgb, 255:red, 0; green, 0; blue, 0 }  ][line width=0.08]  [draw opacity=0] (5.36,-2.57) -- (0,0) -- (5.36,2.57) -- cycle    ;
		\draw   (37.54,192.53) -- (118.86,192.53) -- (118.86,214.28) -- (37.54,214.28) -- cycle ;
		\draw    (75.63,183.25) -- (75.63,189.5) ;
		\draw [shift={(75.63,192.5)}, rotate = 270] [fill={rgb, 255:red, 0; green, 0; blue, 0 }  ][line width=0.08]  [draw opacity=0] (5.36,-2.57) -- (0,0) -- (5.36,2.57) -- cycle    ;
		\draw    (75.63,214.25) -- (75.63,220.5) ;
		\draw [shift={(75.63,223.5)}, rotate = 270] [fill={rgb, 255:red, 0; green, 0; blue, 0 }  ][line width=0.08]  [draw opacity=0] (5.36,-2.57) -- (0,0) -- (5.36,2.57) -- cycle    ;
		\draw   (37.54,223.53) -- (118.86,223.53) -- (118.86,245.28) -- (37.54,245.28) -- cycle ;
		\draw    (119,234) -- (130,234) -- (130,140) -- (122,140) ;
		\draw [shift={(119,140)}, rotate = 360] [fill={rgb, 255:red, 0; green, 0; blue, 0 }  ][line width=0.08]  [draw opacity=0] (5.36,-2.57) -- (0,0) -- (5.36,2.57) -- cycle    ;

		\draw (119.75,49.5) node [anchor=west] [inner sep=0.75pt]   [align=left] {Training set generation using FFD for various basis shapes};
		\draw (119,81.5) node [anchor=west] [inner sep=0.75pt]   [align=left] {Latent space construction by training auto-decoder};
		\draw (147.97,139.25) node [anchor=west] [inner sep=0.75pt]   [align=left] {Shape generator using neural networks};
		\draw (148.13,171.75) node [anchor=west] [inner sep=0.75pt]   [align=left] {Finite elements flow solution};
		\draw (148.09,201.57) node [anchor=west] [inner sep=0.75pt]   [align=left] {Objective computation (particle tracking emulator)};
		\draw (148.02,233.5) node [anchor=west] [inner sep=0.75pt]   [align=left] {Shape parameter update by optimization algorithm};
		\draw (71.33,50.25) node   [align=left] {BS-FFD};
		\draw (71.17,81.73) node   [align=left] {ADAM};
		\draw (79.27,141.53) node   [align=left] {DeepSDF};
		\draw (76.22,172.73) node   [align=left] {FEM};
		\draw (74.47,203.23) node   [align=left] {OBJ};
		\draw (75.45,234.73) node   [align=left] {OPT};
		\draw (15.12,65.68) node  [rotate=-270] [align=left] {offline};
		\draw (14.7,185.47) node  [rotate=-270] [align=left] {online};

	\end{tikzpicture}
	\caption{Pipeline with building blocks of the proposed computational framework.
		The process is split into two parts: A one-time computationally intensive training part and the actual optimization, including the quick filter evaluation.
		To create a training set, \gls{ffd} is applied to a set of basis shapes.
		Subsequently, we train the network using the ADAM optimizer, which concludes the \textit{offline} phase.
		During optimization (i.e., the \textit{online} phase), first, a new shape is created from the neural net.
		Then, a new computational mesh is created around this shape, and based on FEM simulations the new design's mixing is assessed.
		Depending on the objective value, the optimization loop is re-initiated using altered latent variables.
		Building blocks that are modified compared to the general, geometry-kernel-based approach (cf. Fig.~\ref{fig:ffd-framework}) are highlighted in blue.}
	\label{fig:pipeline}
\end{figure}
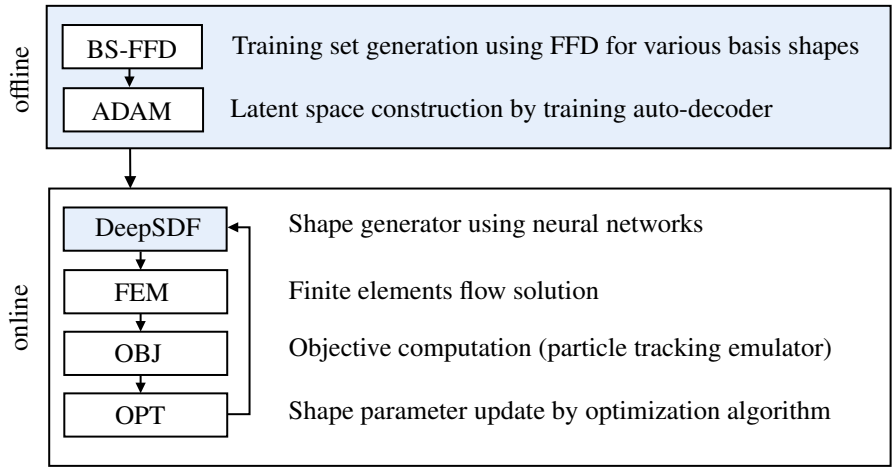

The second algorithm utilized in this work is the single-objective genetic algorithm (SOGA) introduced (as its multi-objective variant) in the JEGA package \cite{Eddy_jega}.
As it belongs to the class of \textit{genetic} algorithms, it solves optimization problems by recreating biological evolution.
Therefore, each optimization run consists of numerous samples referred to as the \textit{population}.
Members of the population are paired and recombined in such ways that the \textit{fitness} (i.e., the objective value) is successively improved.
Regarding its application in this work, it is especially noteworthy that the recreation of evolution includes a \textit{mutation} step, which modifies or re-initializes design variables randomly.
The added randomness allows the algorithm to escape locally convex regions of the design space.
Such evolutionary optimization approaches generally converge slower yielding higher computational costs.
However, they are often able to find better results than non-evolutionary algorithms.
For both DIRECT and SOGA, we rely on the default convergence criterion and a maximum of \num{1000} iterations as a termination criterion.
The complete computational framework is depicted in Fig. \ref{fig:pipeline}.

\section{Numerical results}
\label{sec:results}
This section presents the results obtained using shape parameterizations from neural networks.

Thereby, Sec.~\ref{subsec:latentspacedimension} focuses on the results of the offline phase, i.e., the training of the shape-representing neural network.
In particular, we will discuss the differences in the constructed latent space based on its dimension using the widely used data reduction technique \textit{\gls{tsne}} to visualize the learned, n-dimensional shape parameterization.
In Sec.~\ref{subsec:opt_results} we then present the mixing shapes that could be obtained using our shape optimization approach.

\subsection{Latent space dimension}
\label{subsec:latentspacedimension}

One of the most important choices is the target dimension of the embedding space $l$.
In all established filtering mechanisms like radial basis functions, free-form deformation, \gls{cad}-based approaches, and even mesh-based methods, the practitioner has to balance improved flexibility against the computational demand.
Despite a potentially more compact and dense embedding with neural networks, this is still of relevance and manifests itself in the dimension of the chosen latent space.
Previous works utilized only a very small number of optimization variables.
Elgeti et al. vary between only one and two parameters \cite{Elgeti12}.
Other works by the authors, however, showed that also for six design variables good results are obtained \cite{hube2021}.
To obtain a competitively small number of optimization variables, we  investigate embedding spaces of dimension four, eight, and sixteen respectively, and compare against a free-form-deformation approach using nine variables.\par
Even though the latent space, as discussed in Sec. \ref{subsec:deepGenMods}, in general, obtained lacks an intuitive interpretation, we are still interested in evaluating the quality of the learned embedding space.
We do so in three different ways which we present in the following:
(1) we show a data reduction technique that allows us to visually investigate the latent space;
(2) we apply an interpolation between the latent representation of two training shapes and compare with the expected result;
(3) we apply shape arithmetics, i.e., we isolate a specific modification of a basis shape and impose it onto another basis shape to inspect whether or not features are also recognized by the latent space.\par
\begin{figure}[!h]
	\centering
	\begin{subfigure}[b]{0.32\linewidth}
		\centering
		\includegraphics[width=\linewidth]{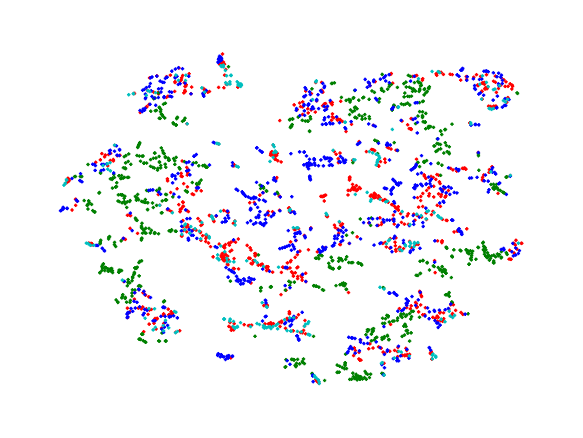}
    	\caption{Training set encoded in 4D latent code}
	\end{subfigure}
	\hfill
	\begin{subfigure}[b]{0.32\linewidth}
		\centering
		\includegraphics[width=\linewidth]{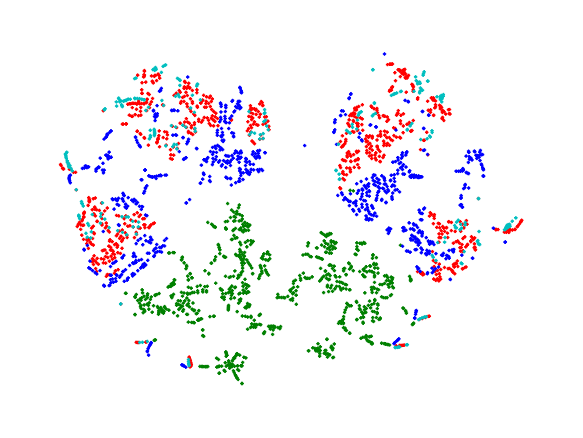}
	    \caption{Training set encoded in 8D latent code}
	\end{subfigure}
	\hfill
	\begin{subfigure}[b]{0.32\linewidth}
		\centering
		\includegraphics[width=\linewidth]{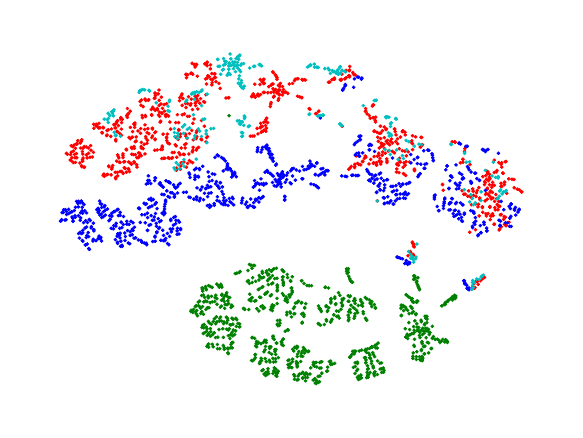}
	    \caption{Training set encoded in 16D latent code}
	\end{subfigure}
    \caption{\gls{tsne} plots obtained using different latent space dimensions. 
    	Increased latent dimension resembles in increased classification performance of the neural net.
    	Each color corresponds to one base training shape: \color{ForestGreen} green \color{black} corresponds a triangular base, \color{blue} dark blue \color{black} is the cube, \color{red} red \color{black}  is the hexahedron, and \color{cyan} light blue \color{black} a tesselated version of the cylinder.}
	\label{fig:tsne_example}
\end{figure}

(1) For the visualization of the high-dimensional latent space, a dimension reduction technique is required.
An intuitive choice might be principal component analysis (PCA), but PCA tries to primarily preserve global structures and thus data points which are far apart in the high-dimensional data will also be drawn far apart in the 2D plot.
Conversely, the correlation between similar points is often lost.
This loss of correlation in similar data is problematic since we aim to investigate whether -- from a human's perspective -- similar shapes are represented by similar latent code.
The problem of loss in local correlation is, however, alleviated by \gls{tsne} \cite{Maaten2008}.
Using \gls{tsne}, we plot each training shape's obtained latent code and -- due to the preservation of local similarities -- similar latent code will form clusters in the scatter plot.
These clusters can then be sampled to verify that the latent code clusters resemble similar shapes.
\gls{tsne} plots for all three latent dimensions -- four, eight, and sixteen -- are shown in Fig.~\ref{fig:tsne_example}.
Fig.~\ref{fig:tsne_example} shows how an increased latent dimension leads to increased classification performance of the neural net.
Specifically, the four chosen basis shapes are clustered with their respective modifications more and more densely as the latent dimension increases.
This improved classification performance indicates that the neural net was able to learn the similarities between similar shapes properly for the case of eight and sixteen dimensions.\par
(2) In addition to comparing clusters of similar shapes in physical and latent space, we also investigate how well the latent space is suited to represent shapes that have not been included in the training set.
We do so by \textit{interpolation} between two shapes.
Fig.~\ref{fig:shape_interpolation} shows the obtained results for all three latent spaces.
\def  \subfigwidth {0.13\textwidth}
\begin{figure}[!h]\centering
	\begin{subfigure}{\linewidth}
		\lineskip=0pt
		\includegraphics[width=\subfigwidth]{figs/interpolation/4/image1.png}\hfill
		\includegraphics[width=\subfigwidth]{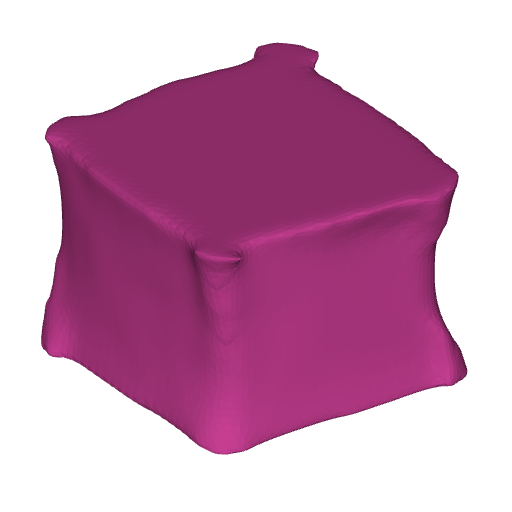}\hfill
		\includegraphics[width=\subfigwidth]{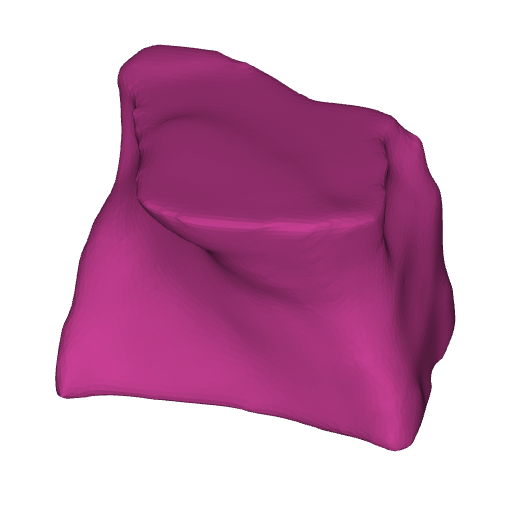}\hfill
		\includegraphics[width=\subfigwidth]{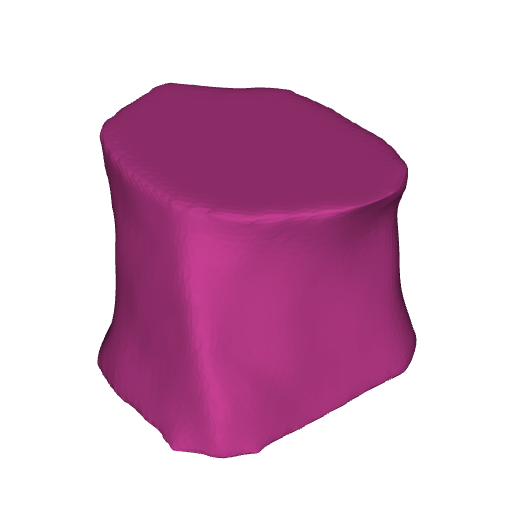}\hfill
		\includegraphics[width=\subfigwidth]{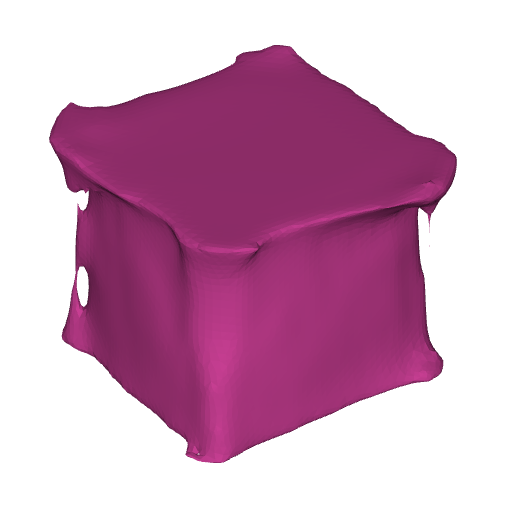}\hfill
		\includegraphics[width=\subfigwidth]{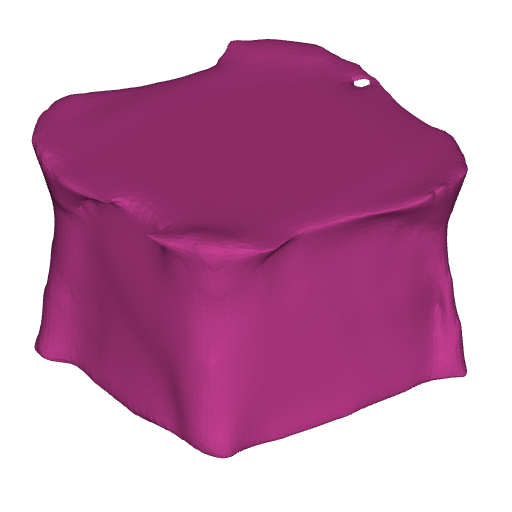}
		\caption{4D shape interpolation revealing artifacts in the reconstructed shapes, i.e. bad quality of the latent representation.}
		\label{fig:interpolation_4D}
	\end{subfigure}%
	\\
	\begin{subfigure}{\linewidth}
		\lineskip=0pt
		\includegraphics[width=\subfigwidth]{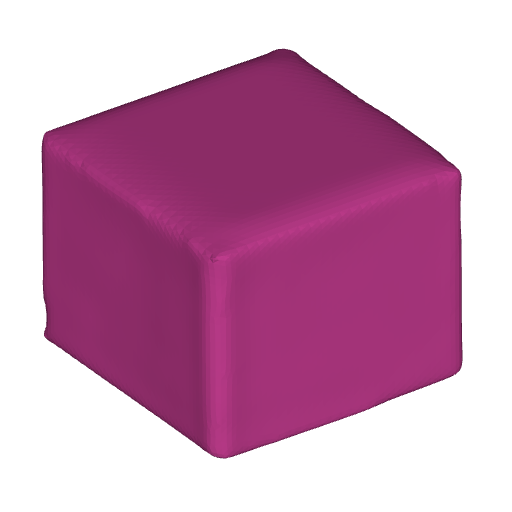}\hfill
		\includegraphics[width=\subfigwidth]{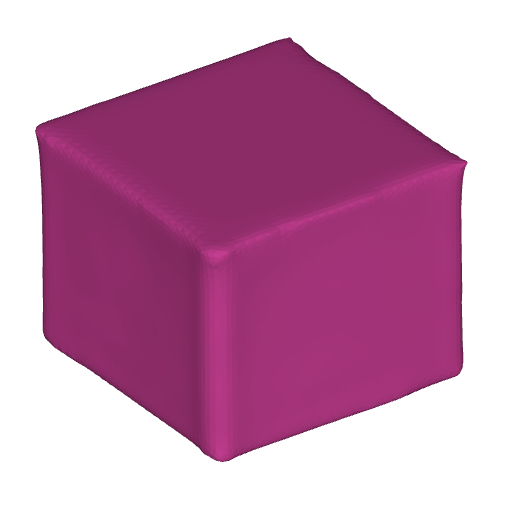}\hfill
		\includegraphics[width=\subfigwidth]{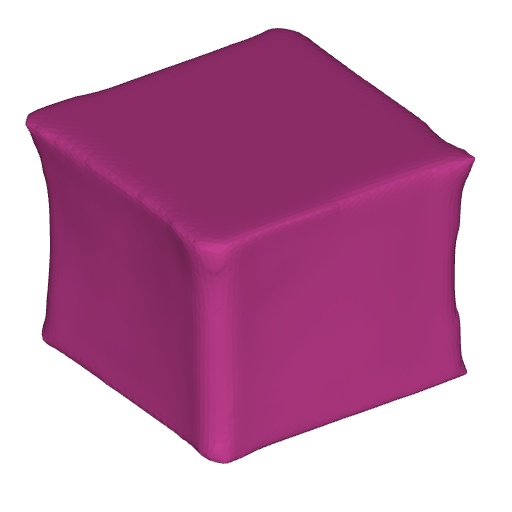}\hfill
		\includegraphics[width=\subfigwidth]{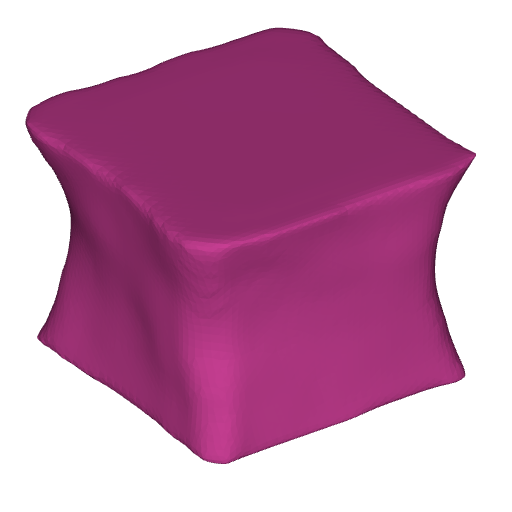}\hfill
		\includegraphics[width=\subfigwidth]{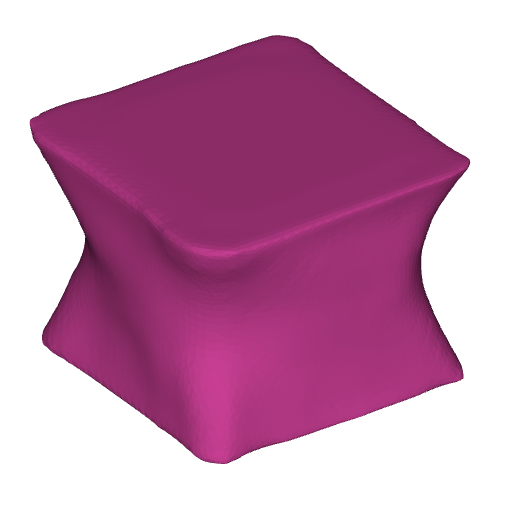}\hfill
		\includegraphics[width=\subfigwidth]{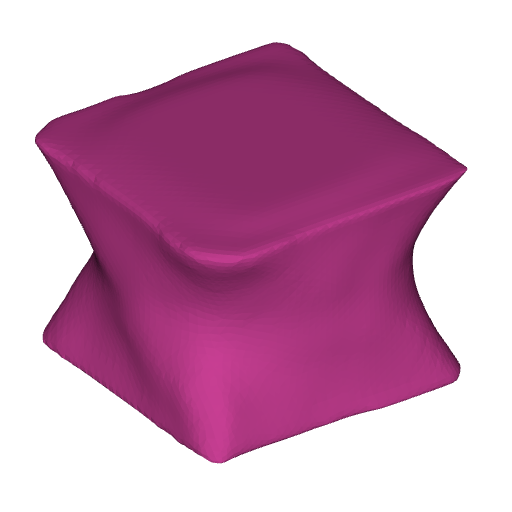}
		\caption{8D shape interpolation with satisfactory results.}
		\label{fig:interpolation_8D}
	\end{subfigure}%
	\\
	\begin{subfigure}{\linewidth}
		\lineskip=0pt
		\includegraphics[width=\subfigwidth]{figs/interpolation/16/image1.png}\hfill
		\includegraphics[width=\subfigwidth]{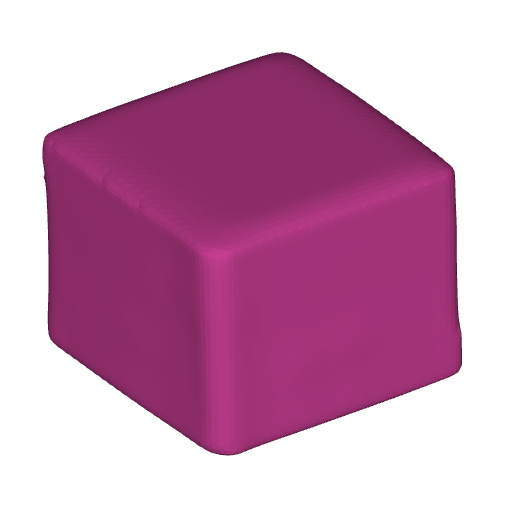}\hfill
		\includegraphics[width=\subfigwidth]{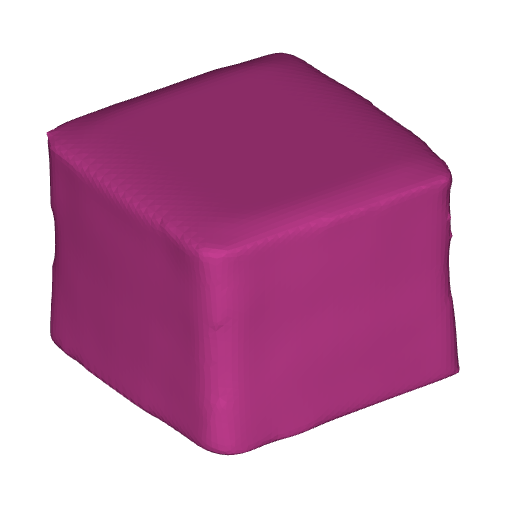}\hfill
		\includegraphics[width=\subfigwidth]{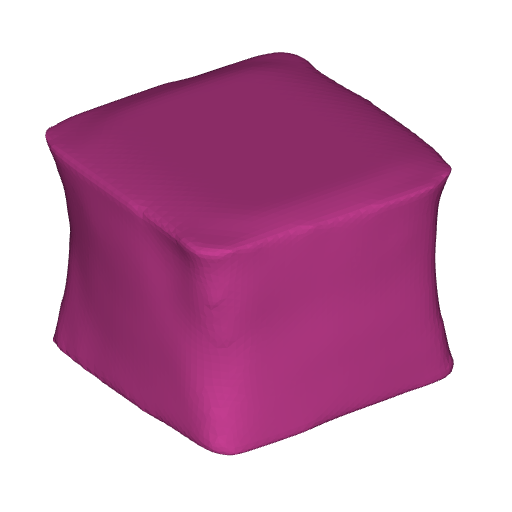}\hfill
		\includegraphics[width=\subfigwidth]{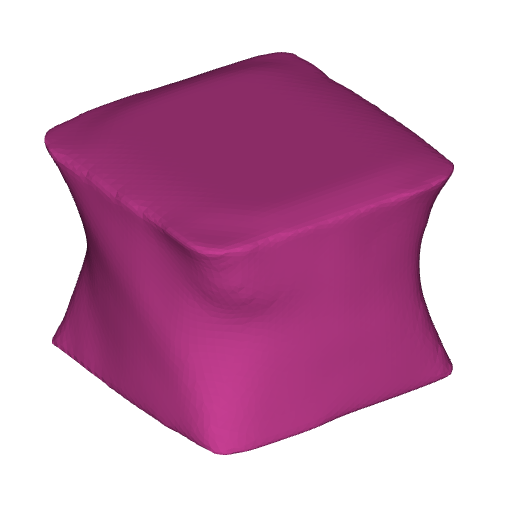}\hfill
		\includegraphics[width=\subfigwidth]{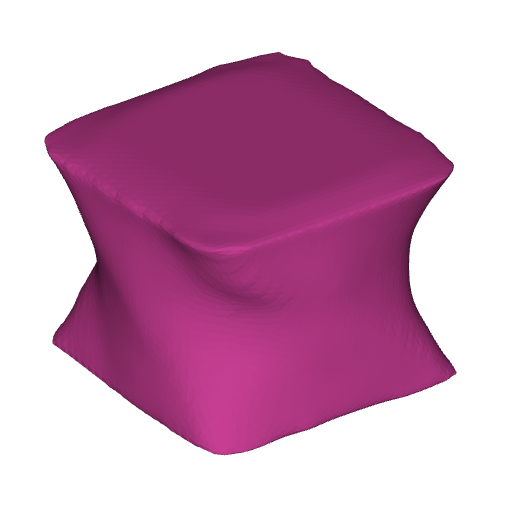}
		\caption{16D shape interpolation, which brings only slight improvement in shape representation compared to eight-dimensional latent representation.}
		\label{fig:interpolation_16D}
	\end{subfigure}
	\caption{Shape interpolation using different latent dimensions.
		An interpolated shape is obtained using $\vekt{z}_{interp}=\vekt{z}_a + \frac{\vekt{z}_b- \vekt{z}_a}{N+1}n$ with $\vekt{z}_a$ and $\vekt{z}_b$ denoting the latent code between shapes $a$ -- here the undeformed cube -- and $b$ -- here the twisted cube.
		With $N=20$, the shown examples represent $n\in [1,3,7,13,17,20]$.}
	\label{fig:shape_interpolation}
\end{figure}
Consistent to the observed lack in classification ability of the four-dimensional latent space, Fig.~\ref{fig:interpolation_4D} shows that interpolation between shapes yields unsatisfactory results.
In particular, shape defects are observed.
This might be a result of the fact that the twisted cube is not at all well represented in the latent space as seen in the rightmost figure.
However, both the eight and the sixteen-dimensional latent space show a visually smooth transition between the regular and the twisted cube shape. \par
\begin{figure}[!h]\centering
	\begin{subfigure}{0.7\linewidth}
		\lineskip=0pt
		\includegraphics[width=0.173\linewidth]{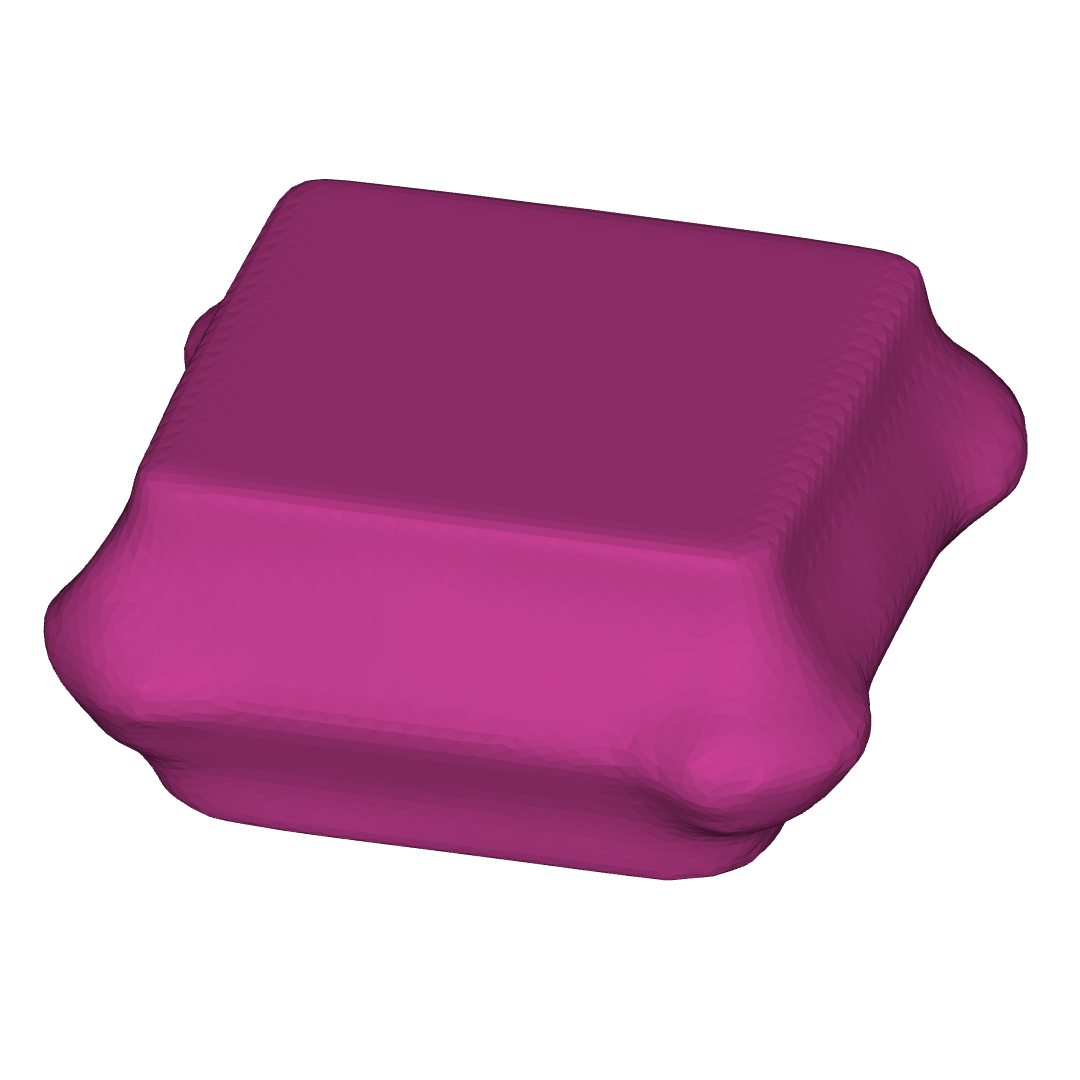}\hfill
		\includegraphics[width=0.173\linewidth]{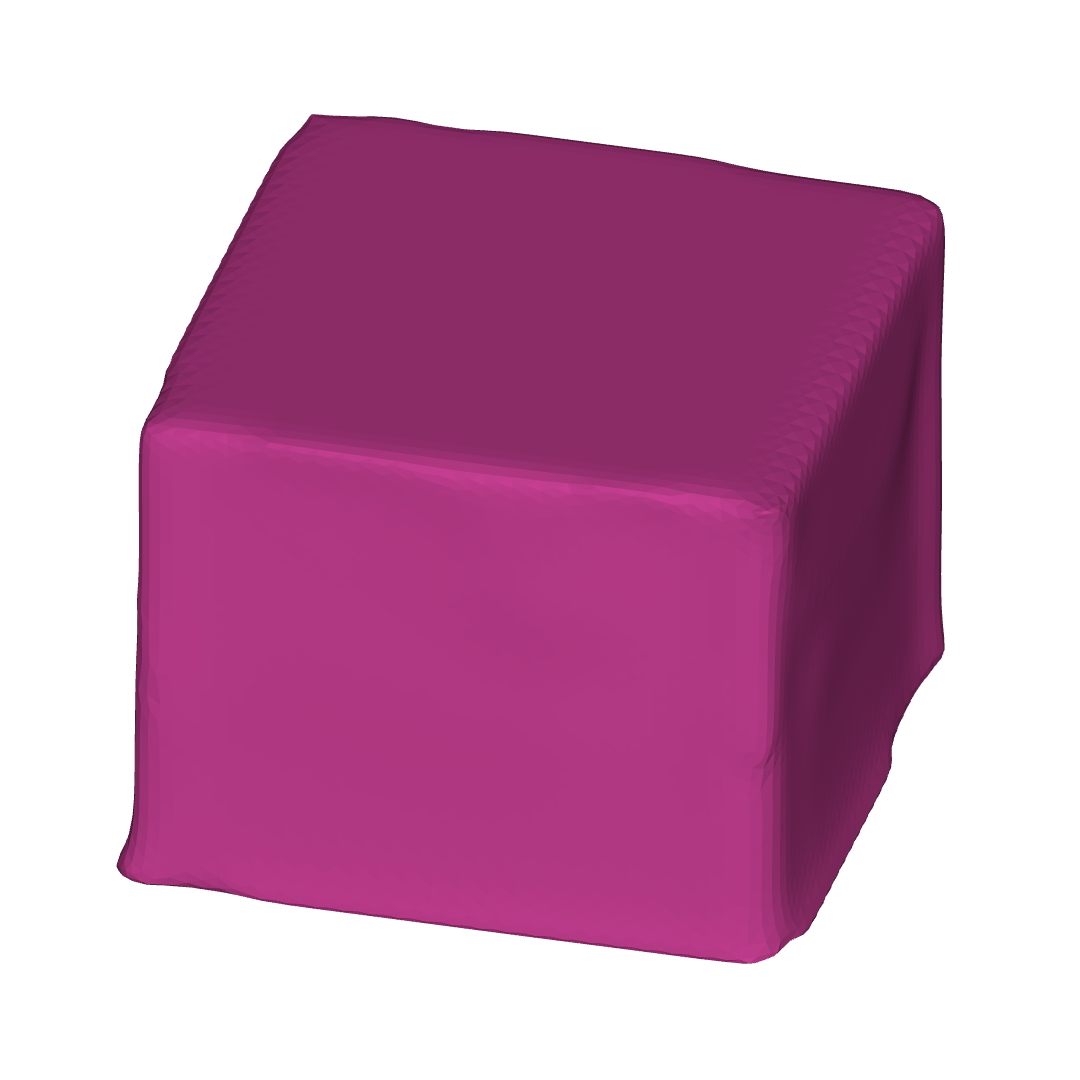}\hfill
		\includegraphics[width=0.173\linewidth]{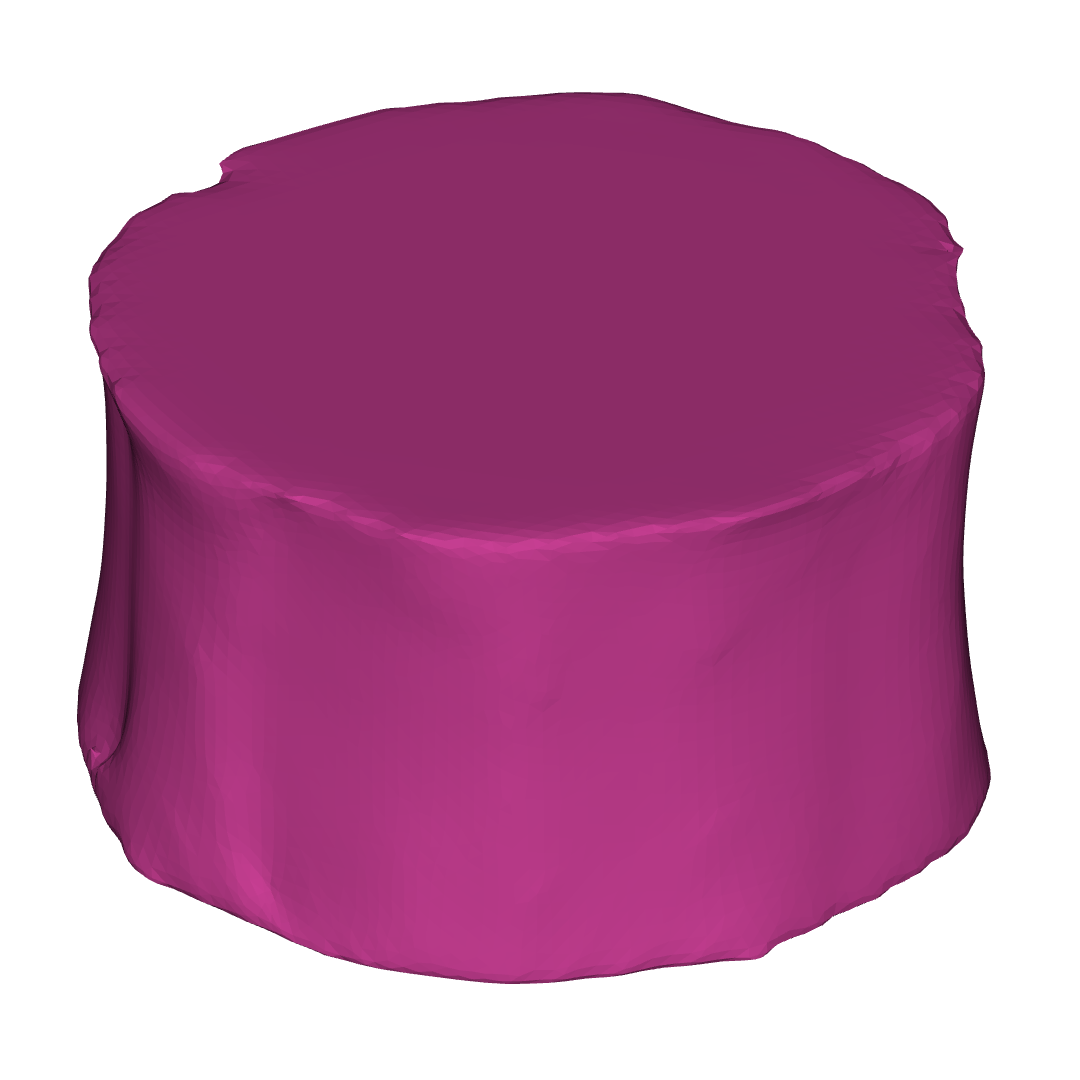}\hfill
		\includegraphics[width=0.173\linewidth]{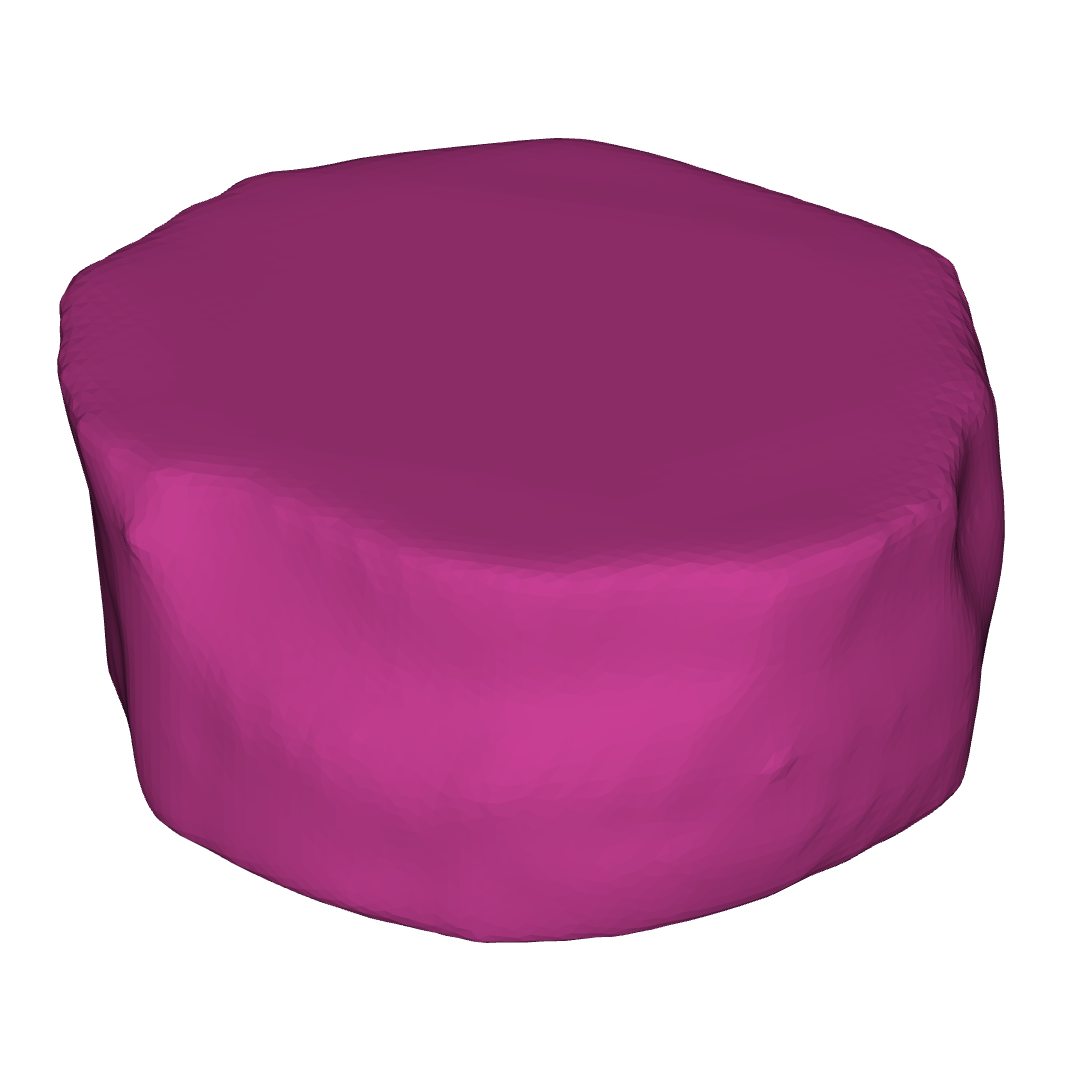}
		\caption{4D shape arithmetics with significant representation errors, especially for the hexahedron and the final shape.}
		\label{fig:arithmetic_4D}
	\end{subfigure}%
	\\
	\begin{subfigure}{0.7\linewidth}
		\lineskip=0pt
		\includegraphics[width=0.173\linewidth]{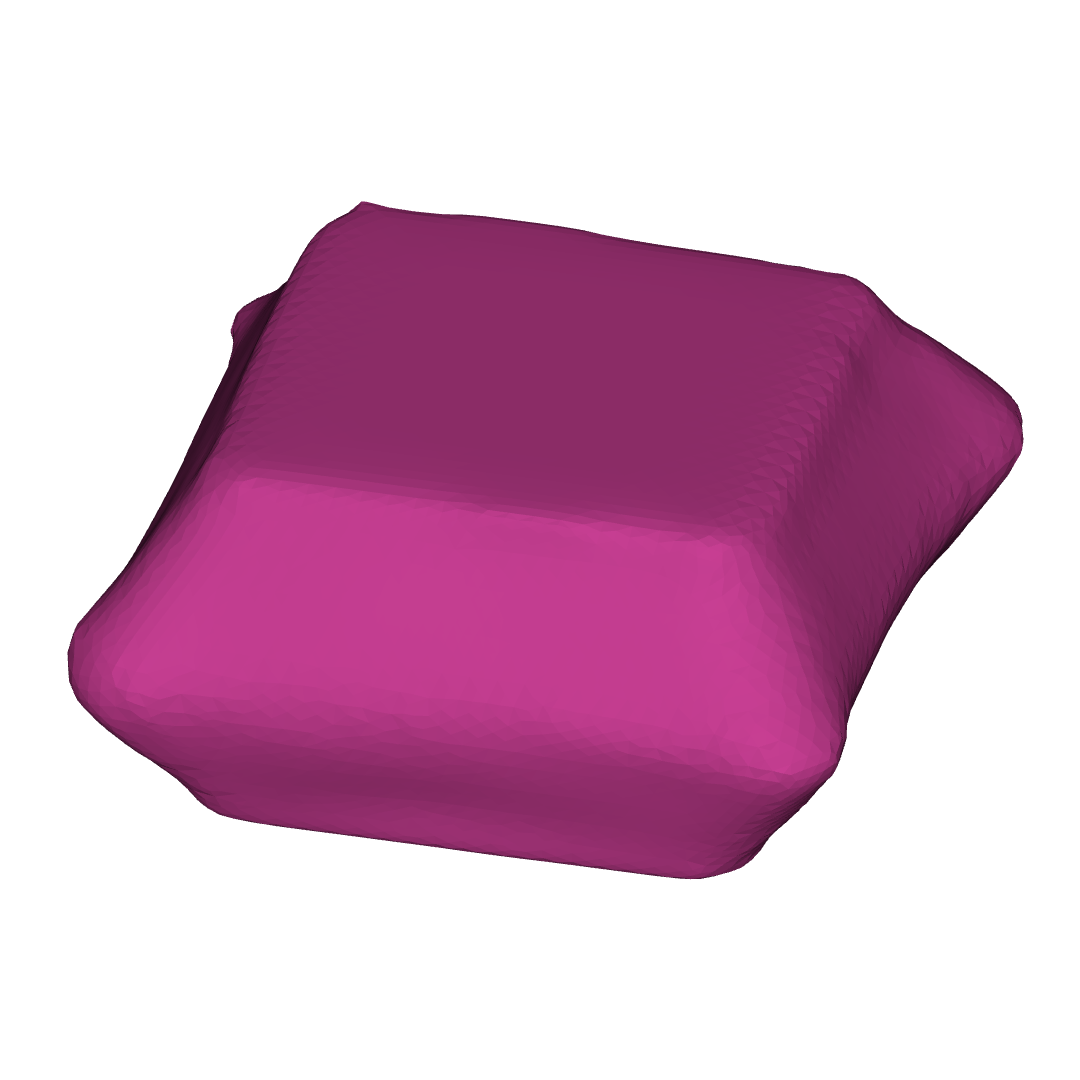}\hfill
		\includegraphics[width=0.173\linewidth]{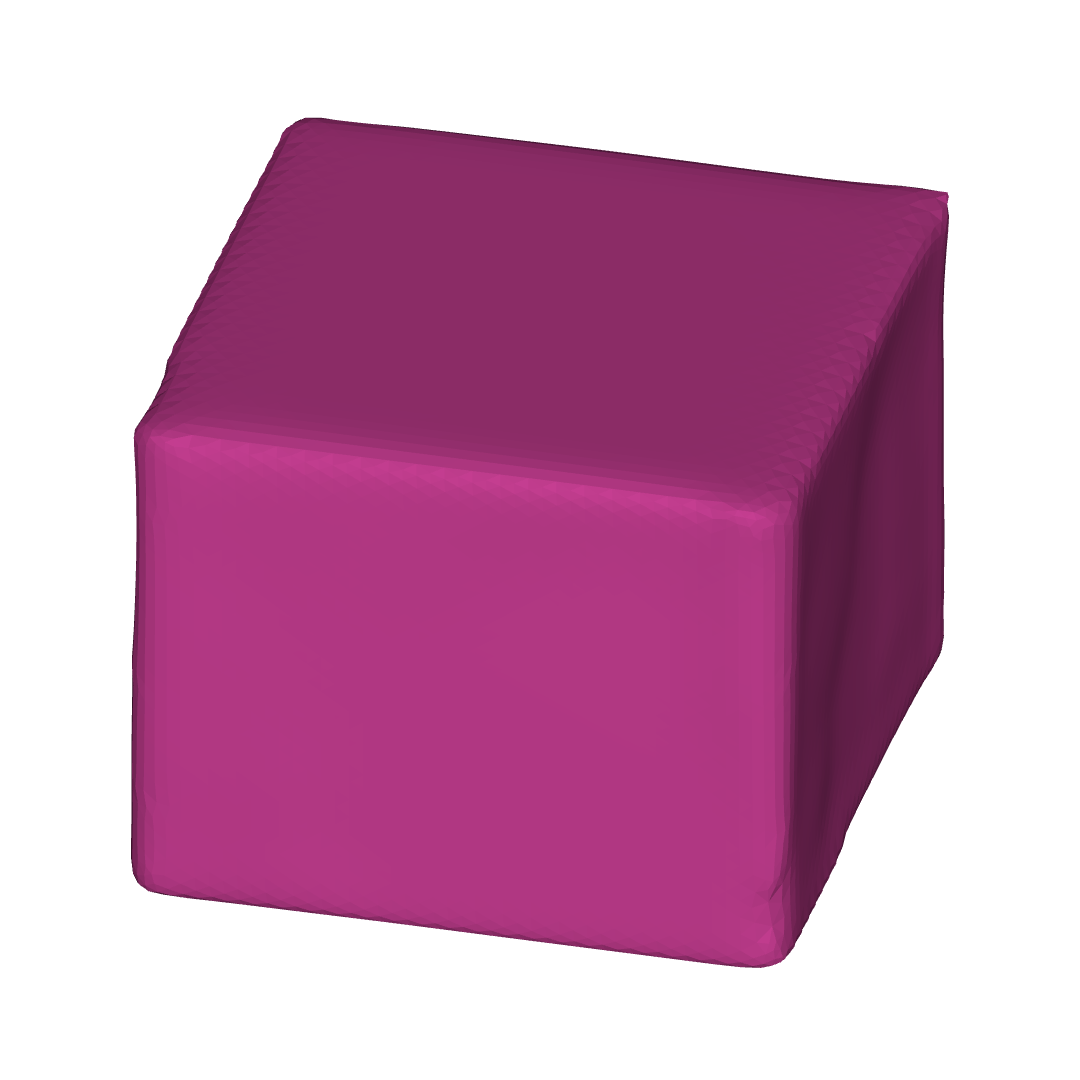}\hfill
		\includegraphics[width=0.173\linewidth]{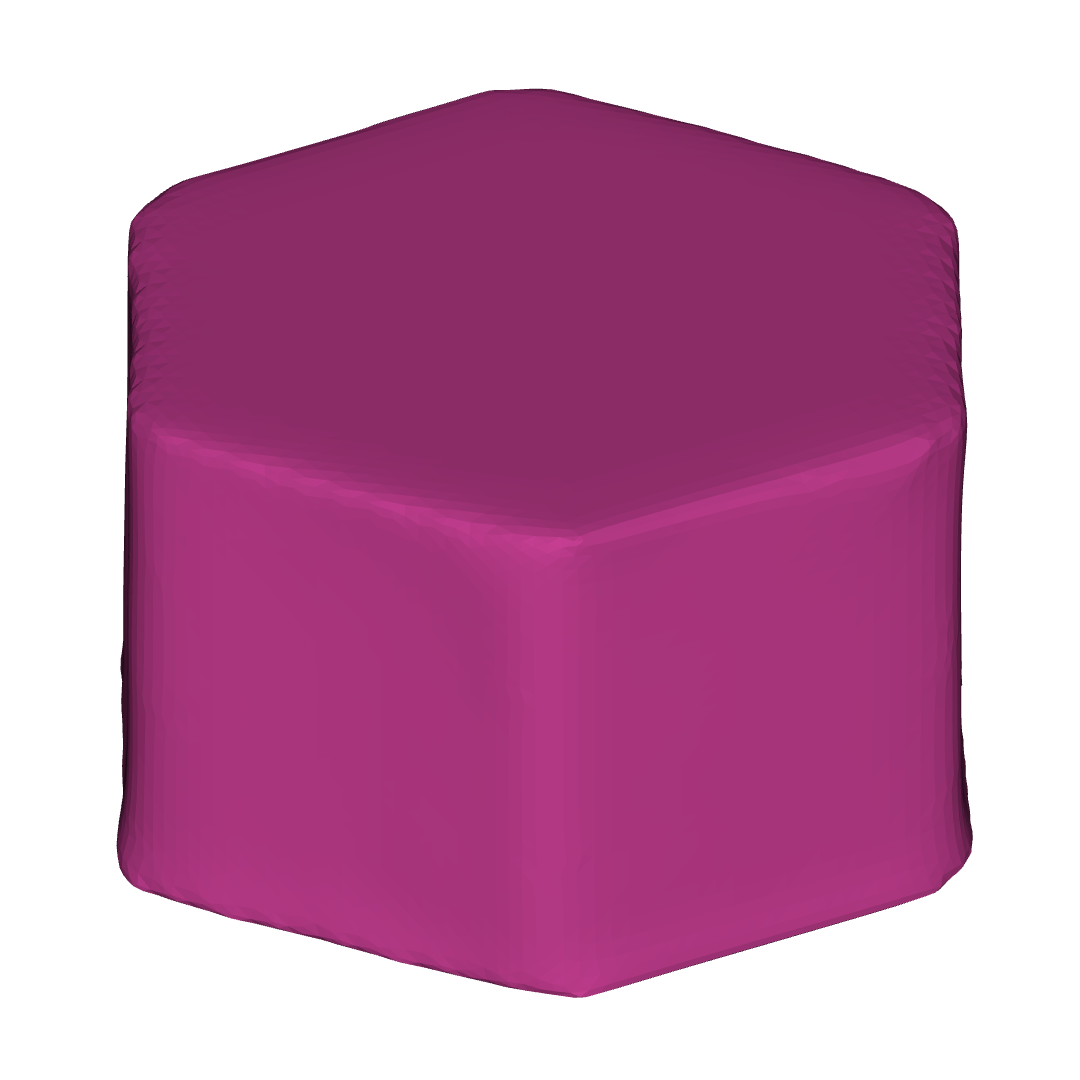}\hfill
		\includegraphics[width=0.173\linewidth]{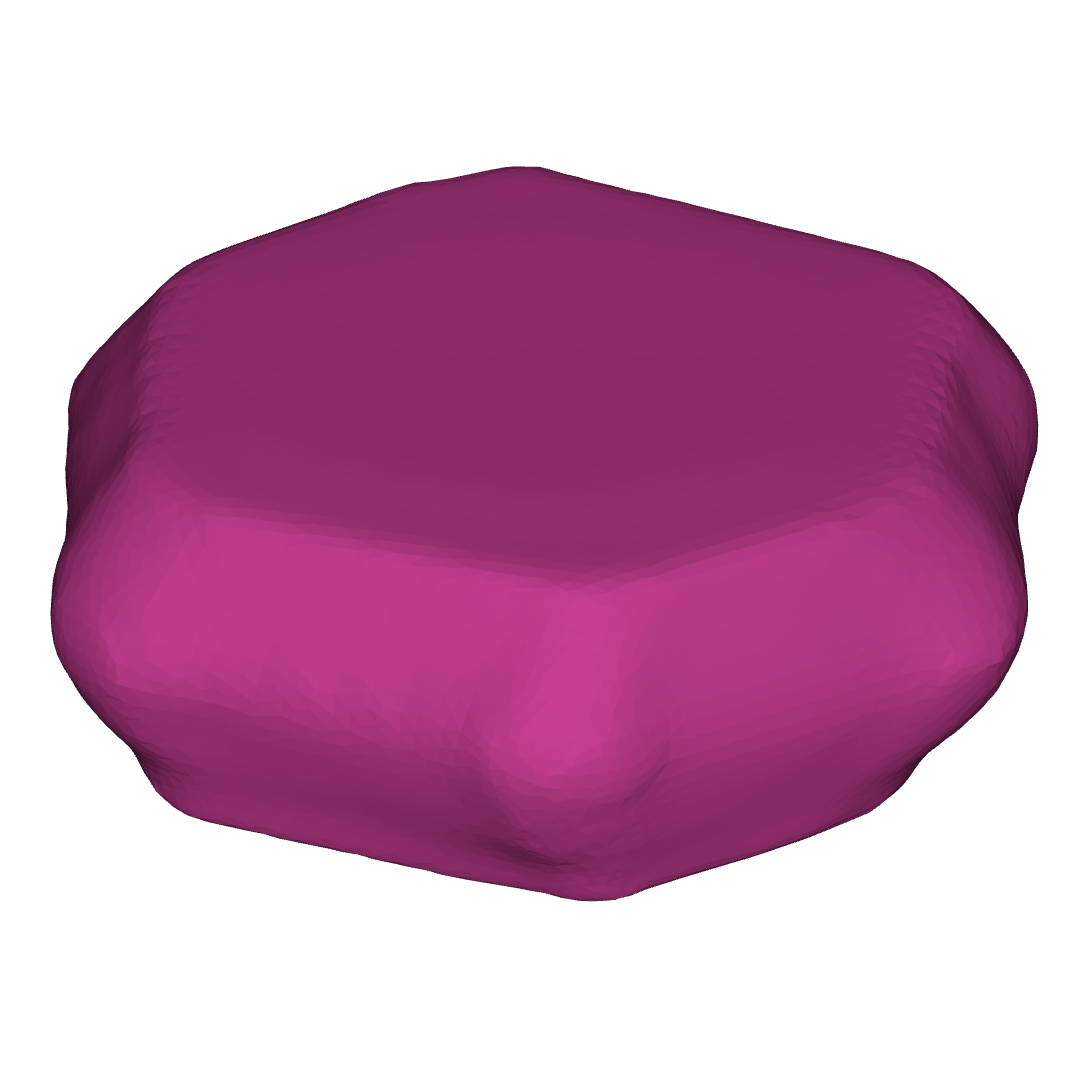}
		\caption{8D shape arithmetic with improved representation compared to 4D latent code but still yielding slightly imprecise results.}
		\label{fig:arithmetic_8D}
	\end{subfigure}%
	\\
	\begin{subfigure}{0.7\linewidth}
		\lineskip=0pt
		\includegraphics[width=0.173\linewidth]{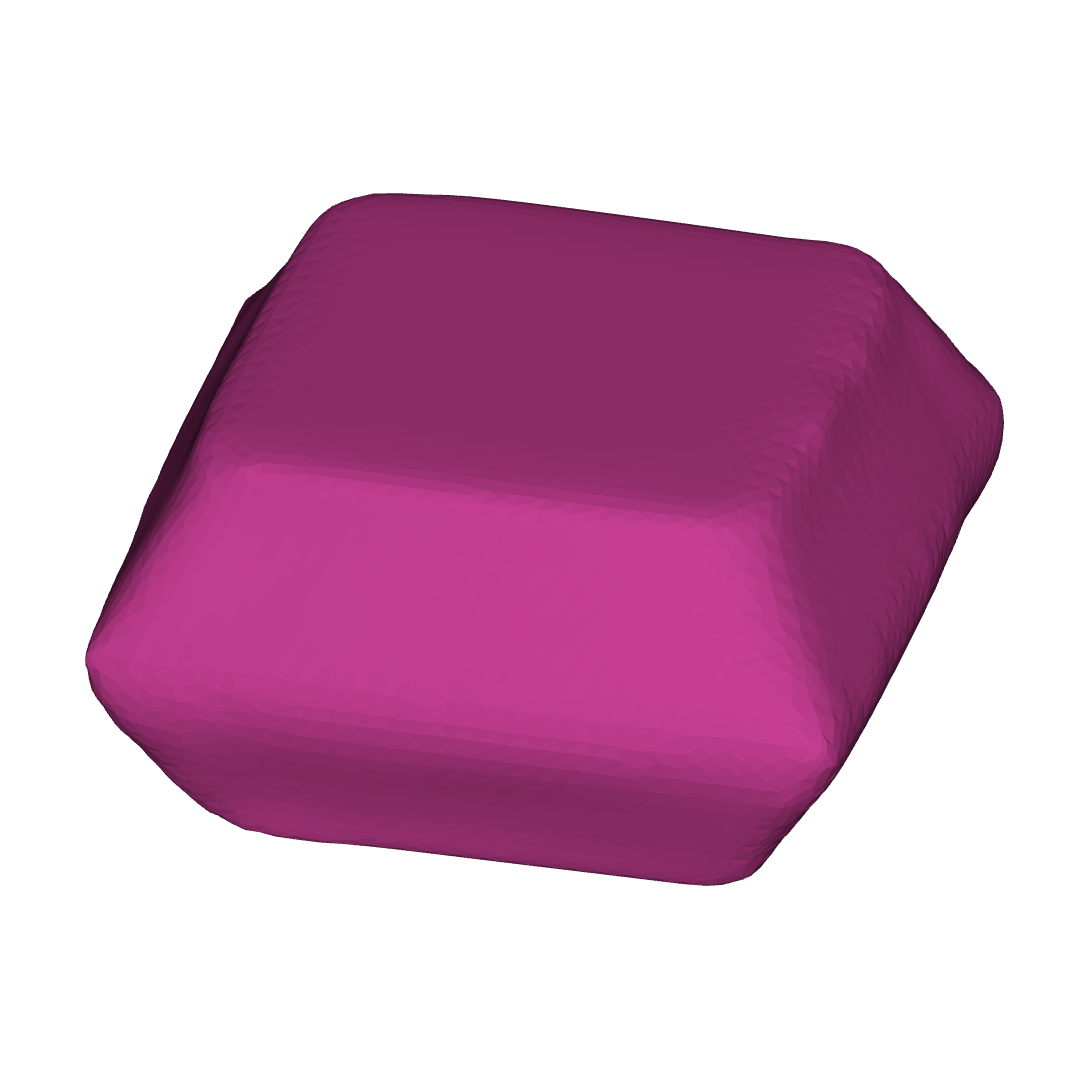}\hfill
		\includegraphics[width=0.173\linewidth]{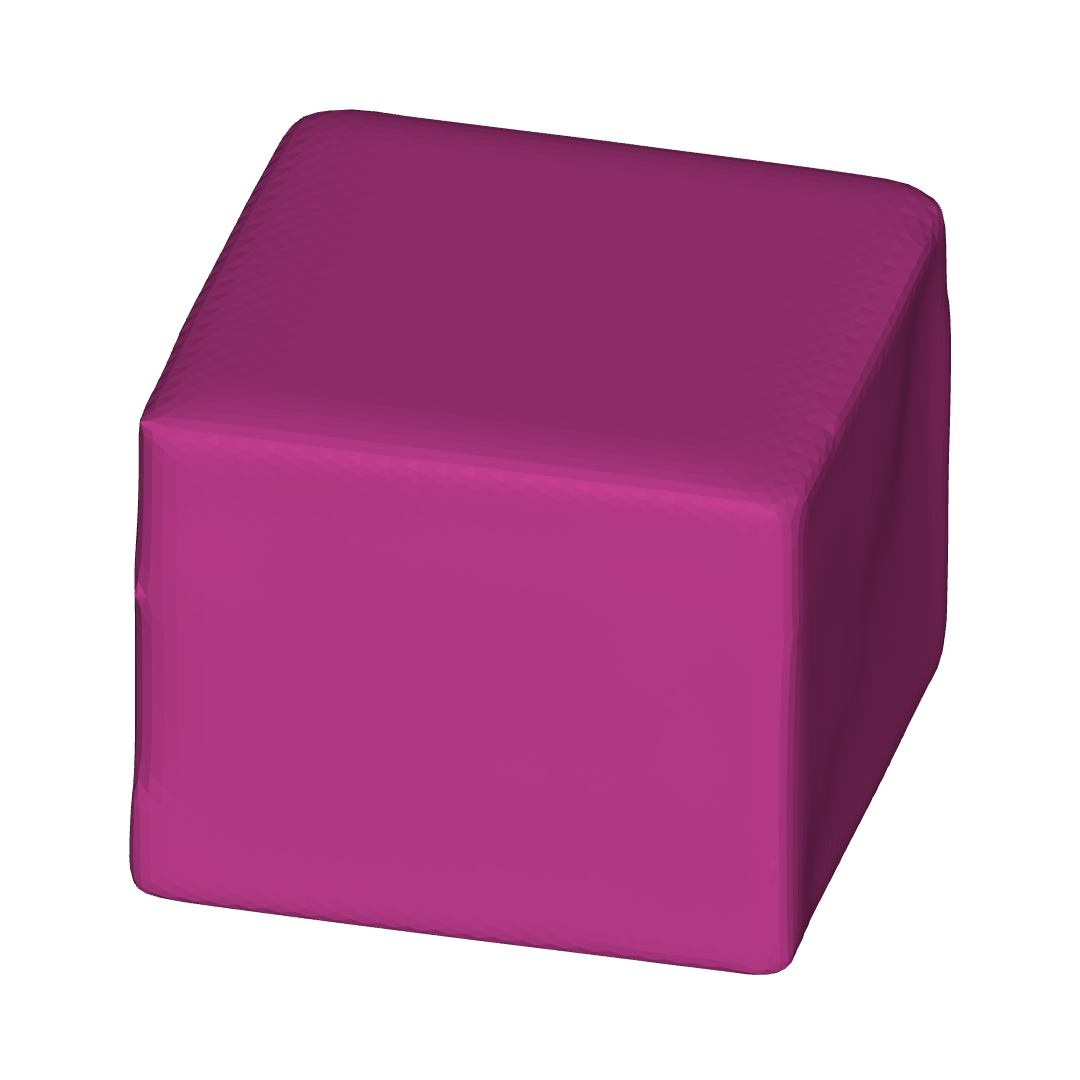}\hfill
		\includegraphics[width=0.173\linewidth]{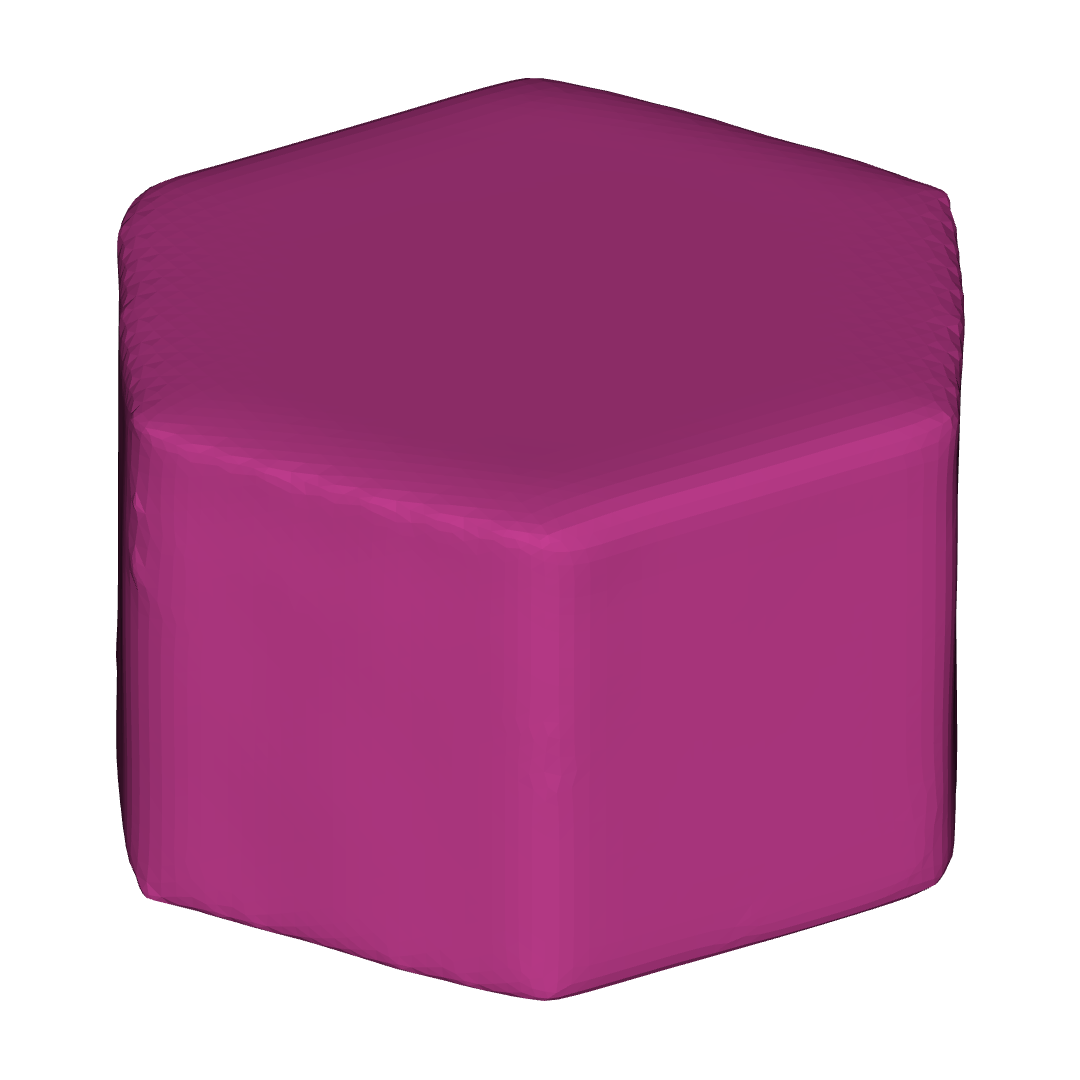}\hfill
		\includegraphics[width=0.173\linewidth]{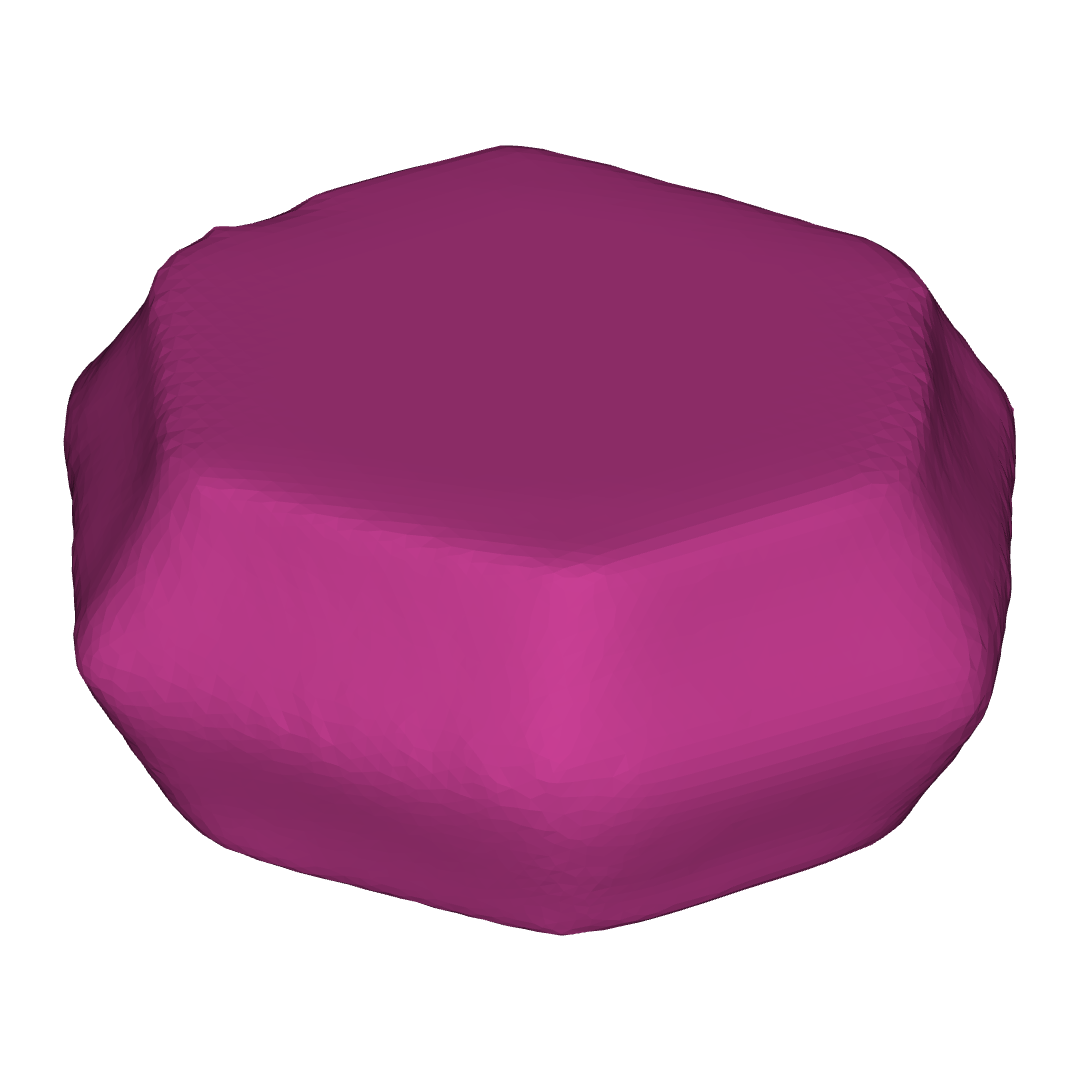}
		\caption{16D shape arithmetic showing perfect resemblance of all training shape and also a clean resulting shape.}
		\label{fig:arithmetic_16D}
	\end{subfigure}
	\caption{Shape arithmetics for different latent dimensions. A linear thickening in the center plane is imposed on a hexagonal base body by evaluation of the latent code as $\vekt{z}_{E4_{thick}}-\vekt{z}_{E4}+\vekt{z}_{E6}$, where $\vekt{z}_{E4_{thick}}$, $\vekt{z}_{E4}$, and $\vekt{z}_{E6}$ denote the latent codes of the thickened cube, the regular cube, and the regular hexahedron respectively.}
	\label{fig:shape_arithmetics}
\end{figure}

(3) The above two analyses investigated the overall classification ability of the neural net and the suitability to represent intermediate shapes.
A final test is given by applying \textit{shape arithmetic}.
Using arithmetic operations applied to the latent code, we extract an exemplary feature -- here a stretching along the center plane -- by taking the component-wise difference of a stretched and a regular cube.
This difference represents center-plane expansion and can then be applied to any other basis shape -- here the undeformed hexahedron.
Fig.~\ref{fig:shape_arithmetics} shows the resulting shapes.
Again, the four-dimensional latent space performs significantly worse since the basis shapes are not represented in detail.
Contrary to the interpolation case, the sixteen-dimensional latent space now shows better results than the eight-dimensional case.\par

All three investigations, \gls{tsne} plots, interpolation and arithmetic indicate that the four-dimensional latent space fails in producing a suitable latent representation.
It should be noted though, that in view of the doubled number of optimization variables, the attainable gains in using sixteen latent variables compared to eight, appear unattractively small.

\subsection{Optimization results}
\label{subsec:opt_results}
To study the effects of the novel shape parameterization technique, we compare configurations that vary in latent space dimensions and optimization algorithms as shown in Tab.~\ref{tab:opt_results}.
Furthermore, we require all generated shapes to have the exact same volume as the undeformed rhombic mixing element utilized in the spline-based optimization (cf. Sec.~\ref{subsec:simulation_model}).
We choose such scaling to avoid convergence towards merely enlarged shapes that yield good objective values but do not deliver helpful insights.
Tab.~\ref{tab:opt_results} lists the obtained results, and Tab.~\ref{tab:opt_iterations} gives insights into the corresponding computational effort.
\begin{table}[!htb]
	\centering
	\begin{tabular}{ |c||c|c|c||c |}
		\hline
		 & 4 & 8 & 16 & \gls{ffd}\\
		\hline\hline
		SOGA  & -0.0726 & -0.0710 & -0.0750 & -- \\
		\hline
		DIRECT & -0.0645 & -0.0738 & -0.0769 & -0.0422 \\
		\hline
	\end{tabular}
	\caption{Different optimization algorithms and latent space dimensions compared by best objective value and contrasted to a nine-dimensional \gls{ffd}.
	Smaller values correspond to better results using the aforementioned objective formulation \ref{eq:objective}.}
	\label{tab:opt_results}
\end{table}
\begin{table}[!htb]
\centering
\begin{tabular}{|c||c|c|c|c|c|c||c|c|}
    \hline
         & \multicolumn{2}{c|}{4}  & \multicolumn{2}{c|}{8}  & \multicolumn{2}{c||}{16}  & \multicolumn{2}{c|}{\gls{ffd}} \\ \hline
    \# Iteration(s) & \multicolumn{1}{c|}{Optimal} & \multicolumn{1}{c|}{Total} & \multicolumn{1}{c|}{Optimal} & \multicolumn{1}{c|}{Total} & \multicolumn{1}{c|}{Optimal} & \multicolumn{1}{c||}{Total} & \multicolumn{1}{c|}{Optimal} & \multicolumn{1}{c|}{Total} \\ \hline\hline
    SOGA & 768 & 1000 & 752 & 1000 & 534 & 1000 & -- & -- \\
    \hline
    DIRECT & 96 & 113 & 129 & 143 & 138 & 149 & 16 & 67 \\
    \hline
    \end{tabular}
	\caption{Different optimization algorithms and latent space dimensions compared by the final iteration count, the obtained objective value, and the number of total iterations; contrasted to a nine-dimensional \gls{ffd}.}
	\label{tab:opt_iterations}
\end{table}
The obtained best shapes are shown in Fig.~\ref{fig:best_opt}.
\def  \figwidth {0.2\linewidth}
\begin{figure}
	\centering
	\begin{subfigure}{\figwidth}
		\centering
		\begin{subfigure}{\linewidth}
			\includegraphics[width=0.85\linewidth]{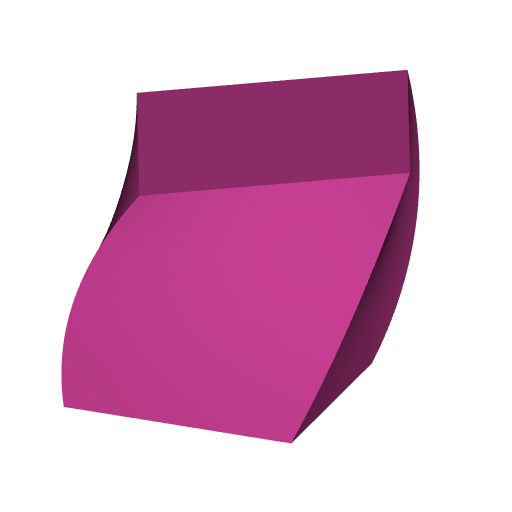}
			\caption{FFD-Direct}
			\label{fig:best_ffd}
		\end{subfigure}
	\end{subfigure}
	\quad
	\hfill
		\begin{subfigure}{\figwidth}
		\centering
		\begin{subfigure}{\linewidth}
			\includegraphics[width=0.85\linewidth]{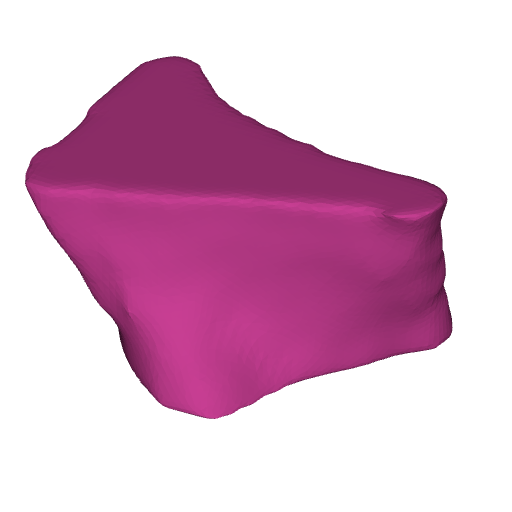}
			\caption{4D-Direct}
			\label{fig:best_opt_4d_direct}
		\end{subfigure}
		\\
		\begin{subfigure}{\linewidth}
			\includegraphics[width=0.85\linewidth]{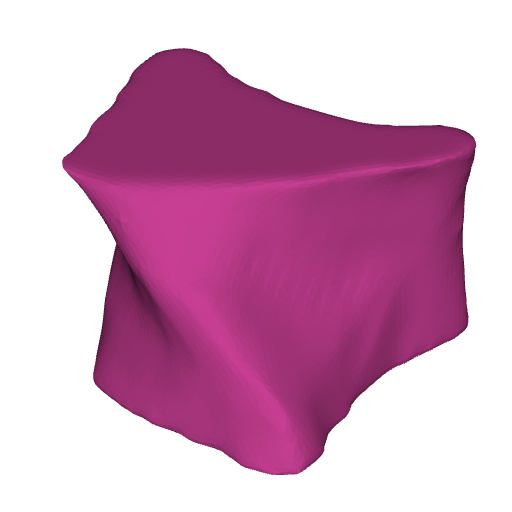}
			\caption{4D-Soga}
			\label{fig:best_opt_4d_soga}
		\end{subfigure}
	\end{subfigure}
	\hfill
		\begin{subfigure}{\figwidth}
		\centering
		\begin{subfigure}{\linewidth}
			\includegraphics[width=0.85\linewidth]{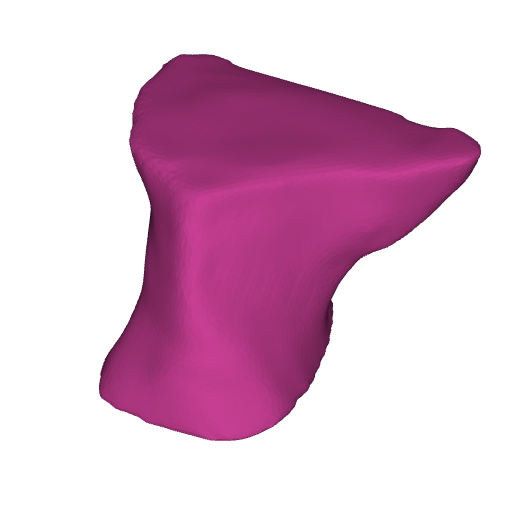}
			\caption{8D-Direct}
			\label{fig:best_opt_8d_direct}
		\end{subfigure}
		\\
		\begin{subfigure}{\linewidth}
			\includegraphics[width=0.85\linewidth]{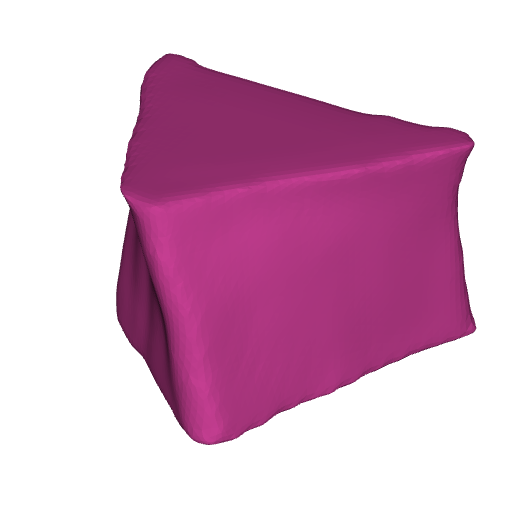}
			\caption{8D-Soga}
			\label{fig:best_opt_8d_soga}
		\end{subfigure}
	\end{subfigure}
	\hfill
		\begin{subfigure}{\figwidth}
		\centering
		\begin{subfigure}{\linewidth}
			\includegraphics[width=0.85\linewidth]{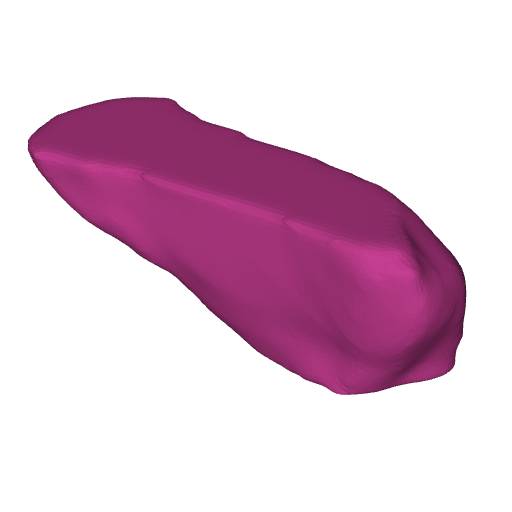}
			\caption{16D-Direct}
			\label{fig:best_opt_16d_direct}
		\end{subfigure}
		\\
		\begin{subfigure}{\linewidth}
			\includegraphics[width=0.85\linewidth]{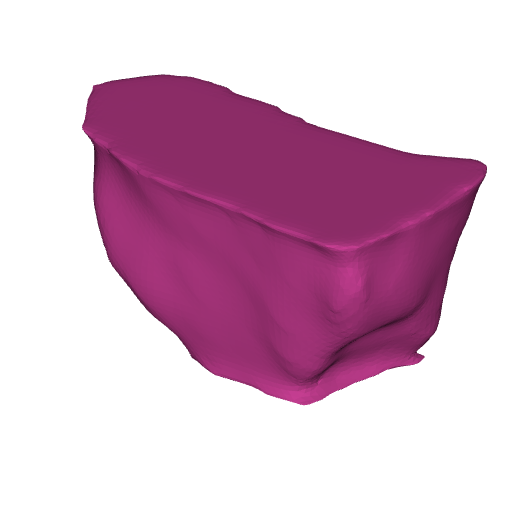}
			\caption{16D-Soga}
			\label{fig:best_opt_16d_soga}
		\end{subfigure}
	\end{subfigure}
	\hfill
	\caption{
		Optimization results obtained for all different latent codes and optimization algorithms compared to an existing \gls{ffd}-based shape optimization.}
	\label{fig:best_opt}
\end{figure}
Comparing the optimized geometries shows interesting results from a plastics processing point of view.
On the one hand, the triangular shape and a mixing element that widens towards the top appear advantageous.
One should note, however, that these deformations do not correspond to a general optimum for plastics engineering but are merely the best possible deformations within the range permitted by the training set.
Choosing an even more diverse training set is expected to yield even further improved shapes.\par
More relevant for this study (with a focus on neural nets as shape parameterizations) is the comparison of convergence, the achieved mixing, and the difference and similarities in the results.
Tab.~\ref{tab:opt_results} shows that for the chosen shape optimization problem, the DIRECT algorithm has no disadvantages compared to SOGA and converges reliably.
Simultaneously, the shape parameterization's dimensionality appears to influence the optimization because the four and eight-dimensional neural networks lead to optimized triangles.
In contrast, the sixteen-dimensional case renders the top-expanded quadrilateral optimal.
Common to all results is a skewed and slightly twisted geometry.\par
A noticeable difference between the spline-based and neural-net-based shape optimization is that the neural-net-based shape parameterization encodes several shapes, of which multiple may mix the melt equally well.
Because of this, from the practitioner's point of view, it does make sense to not only look at the best result but rather compare numerous equally optimal designs and derive design rules from that comparison.
Fig.~\ref{fig:best_opt_16D_soga_compared} shows such a comparison and reveals one advantage of evolutionary algorithms.
\def  \figwidth {0.19\linewidth}
\begin{figure}
	\centering
	\begin{subfigure}{\figwidth}
		\centering
		\includegraphics[width=0.85\linewidth]{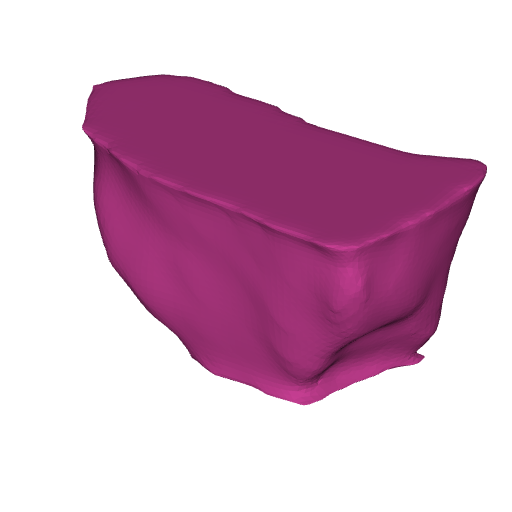}
		\caption{$J=-0.0750$}
		\label{fig:1best_opt_16d_soga}
	\end{subfigure}
	\hfill
	\begin{subfigure}{\figwidth}
		\centering
		\includegraphics[width=0.85\linewidth]{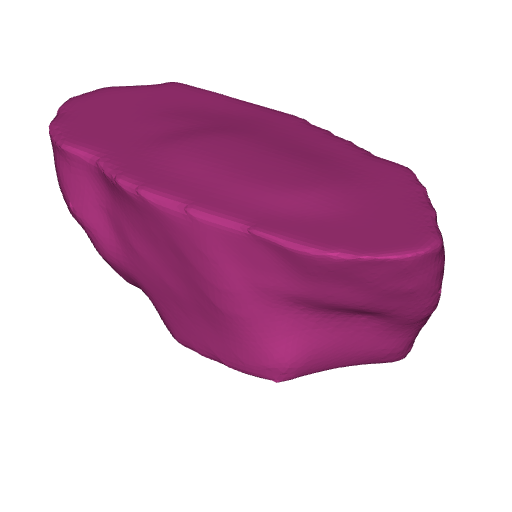}
		\caption{$J=-0.0712$}
		\label{fig:2best_opt_16d_soga}
	\end{subfigure}
	\hfill
	\begin{subfigure}{\figwidth}
		\centering
		\includegraphics[width=0.85\linewidth]{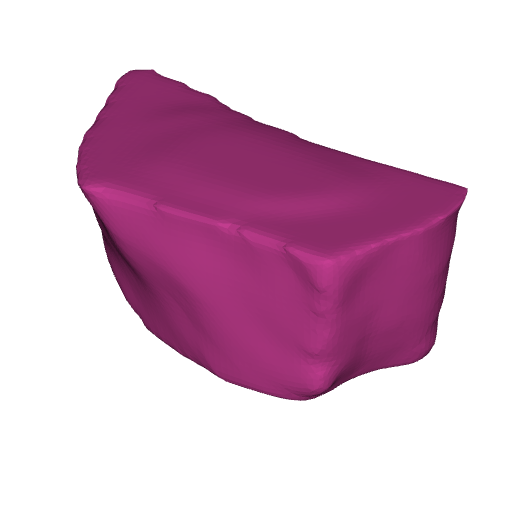}
		\caption{$J=-0.0656$}
		\label{fig:3best_opt_16d_soga}
	\end{subfigure}
	\hfill
	\begin{subfigure}{\figwidth}
		\centering
		\includegraphics[width=0.85\linewidth]{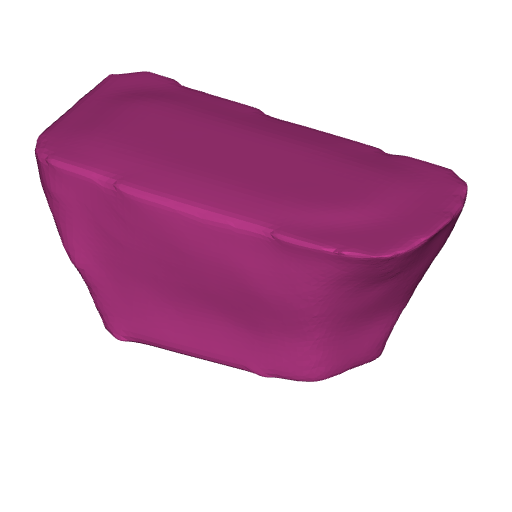}
		\caption{$J=-0.0633$}
		\label{fig:4best_opt_16d_soga}
	\end{subfigure}
	\hfill
	\begin{subfigure}{\figwidth}
		\centering
		\includegraphics[width=0.85\linewidth]{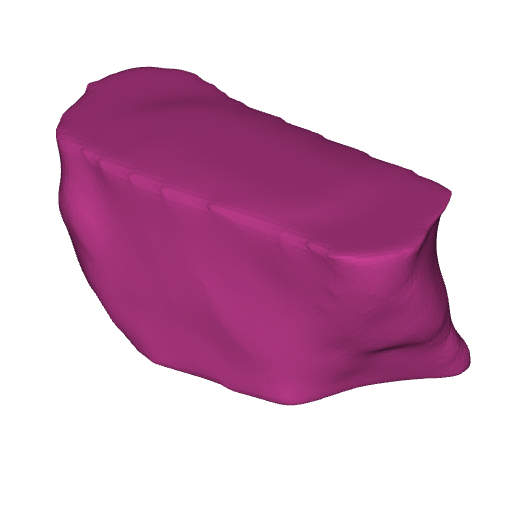}
		\caption{$J=-0.0602$}
		\label{fig:5best_opt_16d_soga}
	\end{subfigure}
	\\
	\begin{subfigure}{\figwidth}
		\centering
		\includegraphics[width=0.85\linewidth]{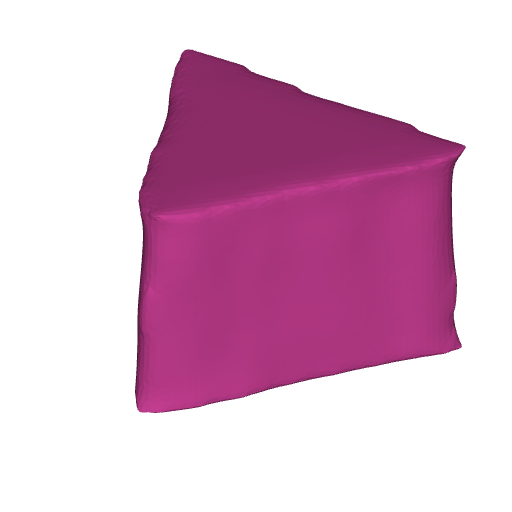}
		\caption{$J=-0.0601$}
		\label{fig:6best_opt_16d_soga}
	\end{subfigure}
	\hfill
	\begin{subfigure}{\figwidth}
		\centering
		\includegraphics[width=0.85\linewidth]{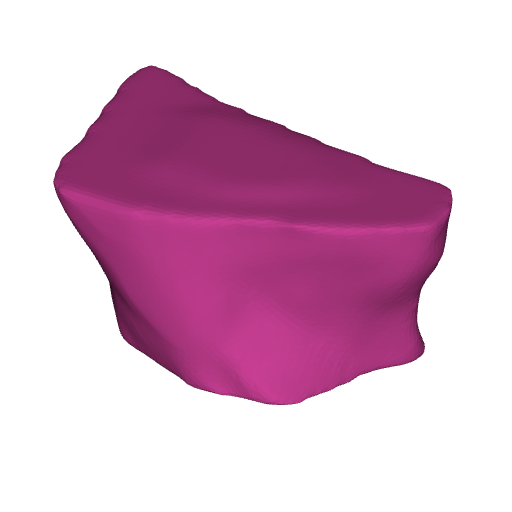}
		\caption{$J=-0.0595$}
		\label{fig:7best_opt_16d_soga}
	\end{subfigure}
	\hfill
	\begin{subfigure}{\figwidth}
		\centering
		\includegraphics[width=0.85\linewidth]{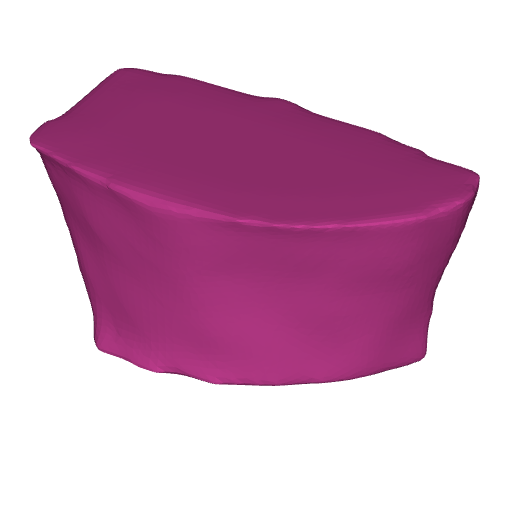}
		\caption{$J=-0.0592$}
		\label{fig:8best_opt_16d_soga}
	\end{subfigure}
	\hfill
	\begin{subfigure}{\figwidth}
		\centering
		\includegraphics[width=0.85\linewidth]{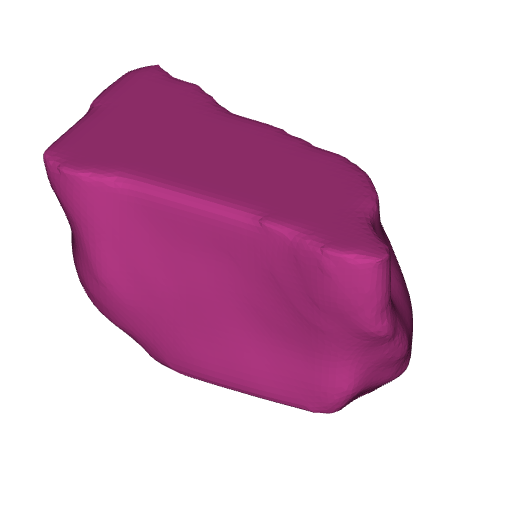}
		\caption{$J=-0.0562$}
		\label{fig:9best_opt_16d_soga}
	\end{subfigure}
	\hfill
	\begin{subfigure}{\figwidth}
		\centering
		\includegraphics[width=0.85\linewidth]{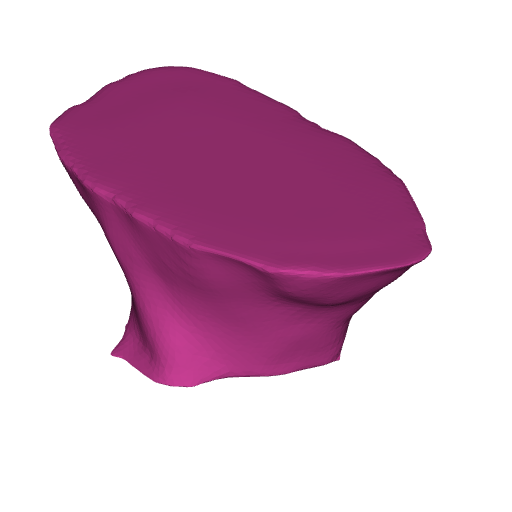}
		\caption{$J=-0.0535$}
		\label{fig:10best_opt_16d_soga}
	\end{subfigure}
	\caption{
		Ten best shapes obtained from 16D SOGA optimization. Except for the 6th-best shape (\subref{fig:8best_opt_16d_soga}), all shapes feature an expanded top, similar orientation, and appear widened in $y$ direction (i.e., perpendicular to the main flow direction).}
	\label{fig:best_opt_16D_soga_compared}
\end{figure}
While the DIRECT algorithm converges locally and, therefore, the ten best designs are geometrically similar, the generative nature of SOGA allows the practitioner to identify possibly equally well-working designs (cf. Figs.~\ref{fig:6best_opt_16d_soga} and \subref{fig:7best_opt_16d_soga}) amongst which the most economical option may be chosen.
Such a choice allows one to account for further restrictions regarding screw cleaning, manufacturability, and others.

\section{Discussion and outlook}
\label{sec:outlook}
In this work, we studied the applicability of generative models as shape parameterizations.
We choose numerical shape optimization of dynamic mixing elements as a use case.
The developed shape parameterization's fundamental principle is to exploit neural nets' ability to construct a dimension reduction onto a feature-dense, low-dimensional latent space.\par
First, the nature of this low dimensional space is studied by \gls{tsne}-plots.
These plots give visual evidence that the generative models create \textit{smooth} shape parameterizations that enable one to use classical, heuristic optimization algorithms.
Comparing genetic to such heuristic algorithms, Tab.~\ref{tab:opt_iterations} reveals that the SOGA algorithm required significantly more iterations (i.e., simulations).
Additionally, Tab.~\ref{tab:opt_results} shows that in the studied examples, this additional computational effort is not reflected proportionally by improved mixing.
One may expect that the SOGA algorithm's random nature may be better suited to explore the hardly interpretable latent space.
However, the results suggest a smoothness of the learned parameterization that renders deterministic methods like DIRECT equally well suited for optimization in the latent space.\par
In addition, to the general applicability of generative models, we study the influence of different latent dimensions.
While the actual optimization results appear pleasing, Figs.~\ref{fig:shape_interpolation} and \ref{fig:shape_arithmetics} suggest that very compressed (i.e., four-dimensional) latent spaces may not be used for optimization purposes.
Analogously, no direct preference between the eight- and sixteen-dimensional results can be drawn from the optimization results.
Similarly, Fig.~\ref{fig:shape_arithmetics} indicates that higher-dimensional latent spaces yield more precise shape encoding, which seems generally preferable.
Since the overall number of iterations until convergence of the optimization problem is comparable, the 16-dimensional parameterization might be chosen over the eight-dimensional variant.\par
As intended, a fundamental improvement over established low-dimensional shape parameterizations is that the new approach covers a much broader design area in a single optimization.
Since its fundamental concept is to encode diverse shapes, optimizations lead to numerous, nearly equally optimal shapes.
Consequently, this novel approach extends on existing methods in that it allows the practitioner to \textit{derive} design features that enhance mixing most and for a wide range of basis shapes.
Therefore, rather than creating complex shape parameterizations, the crucial step towards optimal design reduces to the creative definition of a training set.\par
Finally, a significant challenge in using neural-net-based shape parameterization is proper size control of the output shapes.
This work implements a volume constraint to avoid simple size maximization of the mixing elements.
However, a reformulated objective, such as penalizing pressure loss, may circumvent such adverse designs.
Alternatively, a scale factor may be added as an additional optimization variable.
Both size control and efficient training set generation may be topics of further studies.\par
Given the presented results, utilizing the feature-rich latent representations and their immense generalization power has a significant potential to improve established industrial designs.

\section{Acknowledgements}
The German Research Foundation (DFG) fundings under the DFG grant "Automated design and optimization of dynamic mixing and shear elements for single-screw extruders” and priority program 2231 “Efficient cooling, lubrication and transportation — coupled mechanical and fluid-dynamical simulation methods for efficient production processes (FLUSIMPRO)" - project number 439919057 are gratefully acknowledged.
Implementation was done on the HPC cluster provided by IT Center at RWTH Aachen.
Simulations were performed with computing resources granted by RWTH Aachen University under projects jara0185 and thes0735.


\bibliography{mybibfile}

\end{document}